\documentclass{iopjournal}

\usepackage[square,sort&compress,numbers]{natbib}
\usepackage{enumitem}
\usepackage{graphicx}
\usepackage{amsmath,amssymb,amsfonts}
\usepackage{siunitx}
\usepackage[english]{babel}
\begin{document}

\articletype{Paper}

\title{Measurements, simulations, and models of the point-spread function of electron-beam lithography}

\author{Nikolaj B. Hougs$^{1,*}$\orcid{0009-0000-2625-1201}, Kristian S. Knudsen$^1$, Marcus Albrechtsen$^1$\orcid{0000-0003-4226-0997}, Taichi Suhara$^2$, Christian A. Rosiek$^1$\orcid{0000-0003-3362-4752}, Søren Stobbe$^{1,3,4}$\orcid{0000-0002-0991-041X}}

\affil{$^1$DTU Electro, Department of Electrical and Photonics Engineering, Technical University of Denmark, Ørsteds Plads 343, Kgs.\ Lyngby, DK-2800, Denmark}

\affil{$^2$ELIONIX INC., 3-7-6 Motoyokoyama-cho Hachioji, Tokyo, 192-0063, Japan}

\affil{$^3$NanoPhoton — Center for Nanophotonics, Technical University of Denmark, Ørsteds Plads 343, Kgs.\ Lyngby, DK-2800, Denmark}

\affil{$^4$Beamfox Technologies ApS, Njalsgade 76, 4., Copenhagen S, DK-2300, Denmark}

\affil{$^*$Author to whom any correspondence should be addressed.}

\email{nibaho@dtu.dk}


\begin{abstract}
When a sample is exposed using electron-beam lithography, the electrons scatter deep and far in the substrate, resulting in unwanted deposition of dose at both the nano- and the microscale. This proximity effect can be mitigated by proximity effect correction provided that accurate and validated models of the point-spread function of the electron scattering are available. Most works so far considered a double-Gaussian model of the electron point-spread function, which is very inaccurate for modern electron-beam writers with high acceleration voltages. We present measurements of the process point-spread function for chemically semi-amplified resist on silicon and indium phosphide substrates using a \SI{150}{kV} electron-beam lithography system. We find that the double-Gaussian model deviates from experiments by up to four orders of magnitude. We propose instead a model comprising the sum of a power-law and a Gaussian, which is in excellent agreement with simulations of the electron scattering obtained by a Monte Carlo method. We apply the power-law plus Gaussian model to quantify the electron scattering and proximity effect correction parameters across material stacks, processing, and voltages from \SI{5}{kV} to \SI{150}{kV}. We find that the power-law term remains remarkably constant, whereas the long-range dose contributions and the clearing dose are significantly affected by the substrate and the acceleration voltage. Combined, the results imply that reliance on the double-Gaussian approximation in proximity-effect correction has become a limiting factor for resolution in fabrication at the nanoscale.
\end{abstract}

\section{Introduction}

Electron-beam lithography (EBL)~\cite{nanolithography_2014, bojko_electron_2011} is central to modern nanofabrication as it is the workhorse for both direct-write prototyping in research and for the production of master masks for deep or extreme ultraviolet lithography~\cite{nanolithography_2014,biswas_advances_2012} and nanoimprint lithography~\cite{chou_nanoimprint_1996,nanolithography_2014, asano_metrology_2017}. 
The serial exposure of EBL is slower in comparison, although faster writing speeds are obtained in shaped-beam systems~\cite{ikeno_electron-beam_2016}, multi-beam systems~\cite{servin_progress_2017}, and low-voltage systems~\cite{greve_optimization_2013}.
For high-resolution lithography, the method of choice is single-shot Gaussian-beam EBL using high acceleration voltages~\cite{mankiewich_measurements_1985,jones_very_1987,tennant_progress_2013}.

\begin{figure}[ht]
    \centering 
    \includegraphics[width=0.45\linewidth]{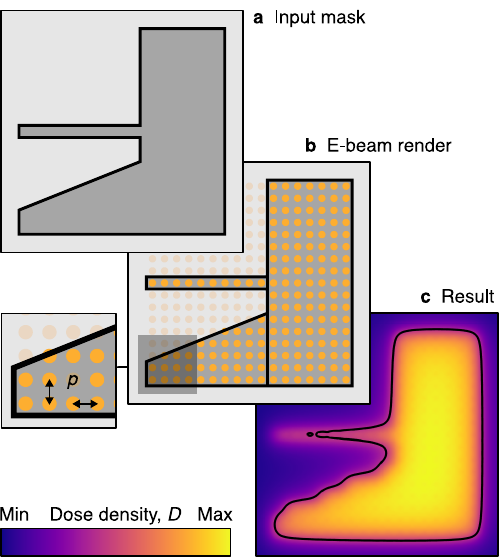}
    \caption{\label{fig:MaskRenderResult}
    Impact of shot filling and proximity effects on electron-beam lithography. (a) An input mask containing the intended pattern is rendered as (b) polygonal exposure primitives. In Gaussian-beam lithography, the polygons are exposed as arrays of shots on a discrete grid (orange dots). (c) The fabricated pattern deviates from the input mask due to three effects: Shot-filling errors, proximity effects, and process effects due to development as well as any subsequent etching and deposition.
    }
\end{figure}

The purpose of EBL is to deposit energy and induce chemical changes in an electron-sensitive resist~\cite{gangnaik_new_2017}, which, after development and subsequent processing such as etching and deposition, results in a mapping of the input mask onto the physical device. However, this mapping is distorted by three effects: Shot-filling errors, the proximity effect, and process effects, as illustrated in Fig.~\ref{fig:MaskRenderResult}. 

Shot-filling errors arise when the input mask is fractured into simple polygons, which, in Gaussian-beam EBL, are further fractured into arrays of shots. In principle, shot-filling errors can be eliminated by using sufficiently small shot pitches, but in practice, limitations in most EBL systems prevent using shot pitches below a nanometer. Although some EBL systems offer features such as shaped beams or additional primitives such as circles, these features cannot fully circumvent shot-filling errors. The only robust solution is to ensure that the input mask consists only of rectangles, whose edges are aligned to the coordinate system of the EBL writer and whose side lengths are integer multiples of the shot pitch~\cite{albrechtsen_nanometer-scale_2022}. 

The proximity effect is much harder to correct for, as it cannot be removed at the design level, and proximity effect correction (PEC) is a hard inverse numerical problem. The issue of the proximity effect and the development of approaches to PEC have therefore been a central element of EBL research since its conception~\cite{eisenmann_proxeccoproximity_1993,mack_electron-beam_2001,parikh_selfconsistent_1978,parikh_corrections_1979,p_kern_novel_1980,owen_proximity_1983,pavkovich_proximity_1986,abe_proximity_1989,owen_methods_1990,murai_fast_1992}. 
The proximity effect depends sensitively on the material stack and exposure parameters, such as the acceleration voltage, $V$, because it is governed by the spatial distribution of deposited energy, \textit{i.e.}, the electron point-spread function (PSF)~\cite{ chang_proximity_1975,rishton_point_1987,stevens_determination_1986,duan_sub10_2010,dai_experiment-based_2011,Manfrinato_Sub-5keV_2011,Winston_modeling_2012,manfrinato_electron-beam_2015}. It can be expanded to a process-specific PSF (PPSF) to include additional effects, such as beam size, resist-development chemistry, and subsequent lithographic steps, \textit{e.g.}, etching~\cite{donnelly_plasma_2013} or deposition~\cite{jensen_chemical_1989,ahmad_deposition_2018}. It is important to stress that even the simplest practical example of EBL includes at least two effects that go beyond the PSF: The finite waist of the electron beam and the development. In practice, this means only the PPSF can be measured.

\begin{figure*}
    \centering
    \includegraphics[width=0.9\linewidth]{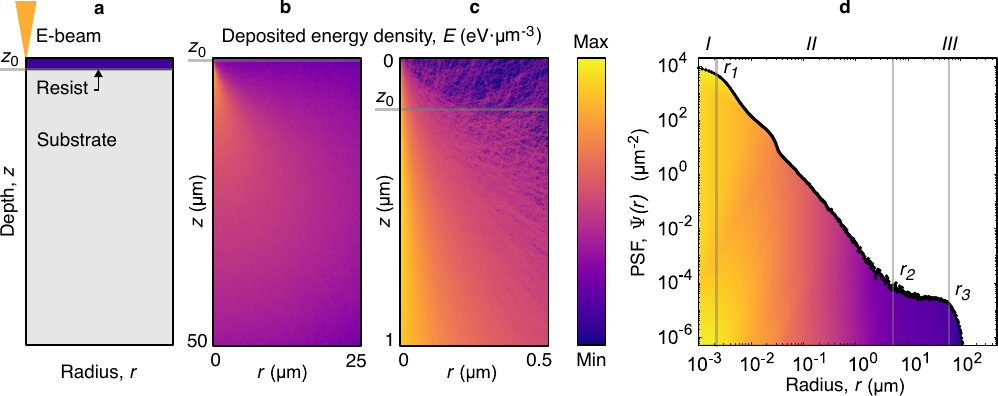}
    \caption{\label{fig:PSFintro}
    The electron point-spread function (PSF) in electron-beam lithography.
    (a) Illustration in cylindrical coordinates of electrons impinging on a stack consisting of resist on a substrate. (b)-(c) Monte Carlo simulation of electron scattering and the corresponding deposited dose distribution from one million \SI{150}{kV} electrons impinging on a resist-silicon stack. (d) When considering only the energy deposition in the resist, the PSF is only a function of the radius, and three distinct regions, $I$-$III$, can be identified as discussed in the main text.
    }
\end{figure*}

In an EBL point exposure, an electron beam impinges on a resist layer on a substrate, see Fig.~\ref{fig:PSFintro}(a). The physics underlying the electron scattering in the resist and the substrate is complex, involving secondary electrons and volume plasmons~\cite{kanaya_penetration_1972,messina_electron_1992,manfrinato_determining_2014,randall_next_2019}, and the proximity effect is therefore commonly modeled using a stochastic Monte Carlo approach~\cite{saitou_monte_1973, hawryluk_energy_1974, adesida_high_1979, mcmillan_simulation_1989, johnson_program_1989, agostinelli_geant4simulation_2003} such as the simulations shown in Fig.~\ref{fig:PSFintro}(b) and (c). The resist is normally much thinner (a few hundred nanometers) than the scattering length (tens of microns), so it is often a good approximation to consider the PSF as being invariant of depth in the resist, in which case the PSF becomes a function of only the radius, $r$, as shown in Fig.~\ref{fig:PSFintro}(d). We identify three distinct regions as indicated in the figure: $I$, for very short ranges, the PSF is relatively flat; $II$, for intermediate ranges, the PSF approximately follows a power-law~\cite{tennant_electron_1988,clauset_power-law_2009}; $III$, for long ranges, the PSF declines approximately with a Gaussian profile, or, as substrate density and electron scattering increases, an increasingly exponential-decay profile~\cite{hawryluk_energy_1974,boere_experimental_1990}.

Ignoring region $II$ and separating dose contributions into incident and backscattering~\cite{saitou_change_1972}, may suggest a double-Gaussian (2G) model which is common in the literature~\cite{chang_proximity_1975,parikh_selfconsistent_1978,parikh_corrections_1979,p_kern_novel_1980,owen_proximity_1983,pavkovich_proximity_1986,abe_proximity_1989,owen_methods_1990,murai_fast_1992}, but this model is an oversimplification. Nevertheless, since only ad-hoc methods exist today for short- and intermediate-range PEC, a common approach to PEC is to use automatic dose modulation for long-range effects~\cite{yu_systematic_2007,bickford_hydrogen_2014,florez_engineering_2022} combined with manual shape correction for short-range effects~\cite{lee_proximity_1991,sewell_control_1978,eunsung_seo_dose_2000,albrechtsen_nanometer-scale_2022}. Although this relies on tedious trial and error for the short-range PEC, it can eventually be an effective strategy. However, this is not an optimal solution, and previous works have improved on the 2G model by modifying the number of Gaussian or exponential terms~\cite{rishton_point_1987, wind_proximity_1989, liu_novel_2008}. Nevertheless, these improved models have not been able to accurately describe PPSFs across a broad range of experimental conditions. 
One reason is that the short-range Gaussian represents the incident electron beam~\cite{chang_proximity_1975}, rather than the electron scattering. Fundamentally, the 2G model is only accurate in the long-range for lighter substrates and in the short-range when the incident beam is very broad. Improvements in EBL technology have moved toward narrower beams and higher $V$, which has gradually made the 2G model a worse approximation of actual PSFs.
Fits with multiple additional Gaussians can be accurate and useful for numerical calculations, but such fits inevitably include many parameters that impede assessments of resolution limits and quantitative comparisons between different processes across different hardware and material platforms. 
More accurate PSF models significantly improve PEC~\cite{liu_impacts_2013} and are thus necessary, especially at high $V$. To this end, recent models of the PSF have included power laws~\cite{albrechtsen_power-law_2019,mao_proximity_2025}, but investigations across higher acceleration voltages and, more importantly, systematic comparisons to experiments were so far missing.

Here we present accurate measurements of the PPSF for chemically semi-amplified resist (CSAR)~\cite{schirmer_chemical_2013} on both silicon and indium phosphide (InP) substrates at \SI{150}{kV} using single-shot exposures. We find excellent agreement with Monte Carlo simulations of the PSF except for distances below 10 nm, which indicates that the PSF and PPSF are identical except at very short distances. We propose an empirical power-law plus Gaussian (PLG) model of the PSF, and we demonstrate a greatly improved agreement with experiment over 10 orders of magnitude of deposited dose, corresponding to 5 orders of magnitude of distances from \SI{1}{nm}-\SI{100}{\micro\meter} for both silicon and InP substrates.
The PLG model is validated to enable quantitative comparisons across exposure voltages $5-\SI{150}{kV}$, which suggest that reliance on 2G-model-based PEC is a limiting factor in modern high-resolution EBL.

\section{Method: Measuring the point-spread function}

Before discussing our experiment, we briefly outline the theory of EBL and the theoretical foundation for extracting the PSF, building in part on Refs.~\cite{chang_proximity_1975,rishton_point_1987}. The purpose of EBL is to map a mask, $M(x,y)$, onto a physical device with the greatest possible fidelity. In Gaussian-beam EBL, the mask is rendered as a shot map
\begin{equation}\label{eq:shot_pitch}
    S(x,y,p) = p^2\sum_{i,j} \delta(x-i\,p) \delta(y-j\,p) M(x,y),
\end{equation}
where $p=p_x=p_y$ is the shot pitch. It is desirable to choose $p$ much smaller than the smallest critical dimension of the mask to minimize shot-filling errors. Each shot is broadened according to the PSF, $\Psi(r)$, which is related to the density of energy deposited in the resist, $E(r)$, as
\begin{equation}\label{eq:PSF-definition-radial}
    \Psi(r) = \frac{E(r)}{2\pi \int E(r) \, r\,dr}.
\end{equation}
The PSF is a probability density with normalization $2\pi \int \Psi(r)  \, r\,dr = 1$. We refer to Appendix~\ref{app:psf_from_monte_carlo} for additional details on extracting the PSF from simulations of $E(r)$. In practice, the energy is often replaced by an effective deposited charge dose density
\begin{equation}\label{eq:charge-to-energy}
    D(r) = k^{-1}\,E(r),
\end{equation}
with $k$ a charge-to-energy conversion factor that is constant for a given process and exposure condition. With these definitions, the deposited dose distribution is given by
\begin{equation}\label{eq:dose_distribution}
    D(x,y) = D_\mathrm{in} \Psi(r) \ast S(x,y,p),
\end{equation}
where $\ast$ denotes the convolution, and $D_\mathrm{in}$ is the impinging charge dose density, which is directly related to the impinging shot charge, $q_\mathrm{in} = p^2 D_\mathrm{in}$. 
Thus, if no PEC is applied, the convolution of $S(x,y,p)$ with $\Psi(r)$ distorts $D(x,y)$. The purpose of PEC is to transform $S(x,y,p)$ and/or apply a spatial variation to $D_\mathrm{in}$, such that $D(x,y)$ becomes as close to the input mask as possible. After development, the resist is fully cleared where
\begin{equation}\label{eq:D0_definition}
    D(x,y) \geq D_0,
\end{equation}
with $D_0$ the clearing dose, which is the minimum dose required to clear the center of a polygon whose size is much larger than any length scales in the PSF. An advantage of this definition is that the clearing dose is pattern-independent, \textit{i.e.}, in the center of a sufficiently large polygon, $D(x,y)$ becomes equal to $D_{\mathrm{in}}$. Note that $D_0$ depends on the material stack, acceleration voltage, and development conditions.

The exposed pattern can only be imaged after development, which implies that an experiment probes the PPSF rather than the PSF. It is useful to consider an isolated shot, where the shot map reduces to $S(x,y,p)=p^2 \delta(x,y)$ and Eq.~(\ref{eq:dose_distribution}) reduces to
\begin{equation}\label{eq:dose_single_shot}
    D(r,p) = p^2 D_\mathrm{in} \Psi(r) = q_\mathrm{in} \Psi(r),
\end{equation}
Combined with Eq.~(\ref{eq:D0_definition}), this ensures that the resist is fully developed out to a radius $r_0$, such that the PPSF can be obtained from a sequence of single-shot exposures with different $q_{\mathrm{in}}$ using
\begin{equation}\label{eq:psf_measurement_theory}
    \Psi(r_{0}) = \frac{D_0}{q_{\mathrm{in}}}.
\end{equation}

\section{Results and discussion}

\subsection{Measuring the clearing dose}

\begin{figure*}[h]
    \centering
    \includegraphics[width=0.85\linewidth]{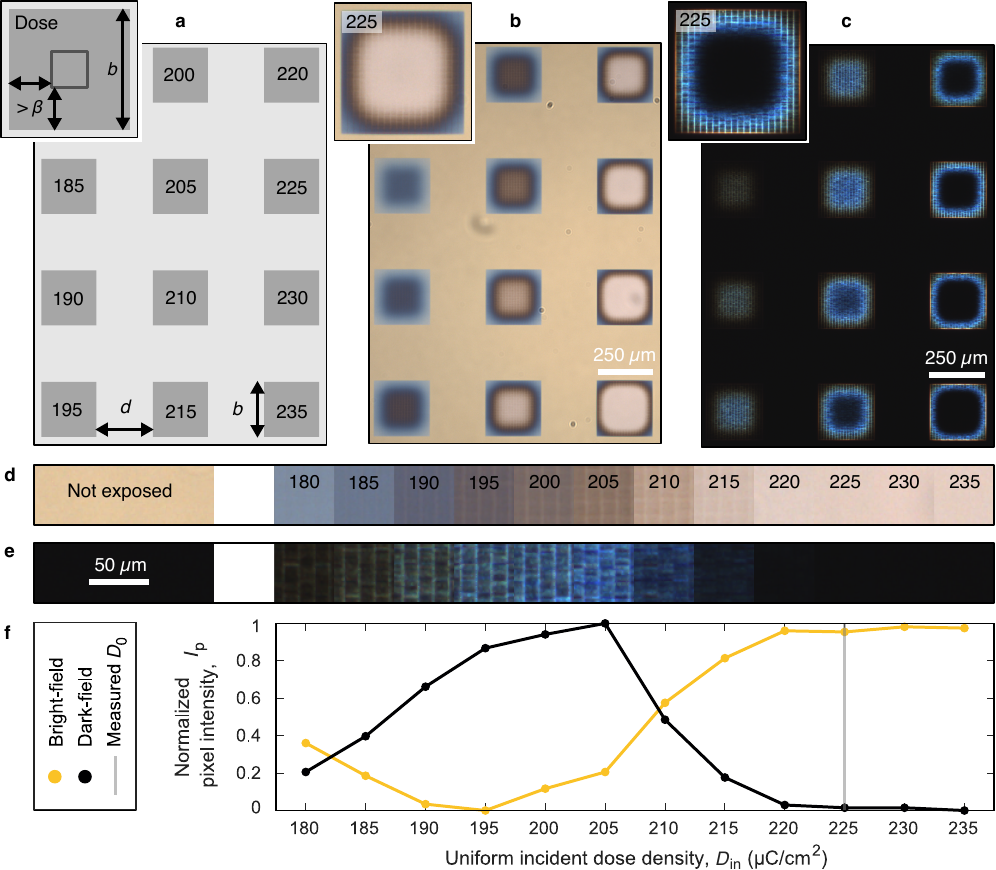}
    \caption{\label{fig:ClearingDose}
    Measurement of the clearing dose, $D_0$, for \SI{150}{kV} exposures on a stack comprising \SI{180}{nm} CSAR on silicon. (a) A mask containing squares with uniform incident doses indicated as the numbers in units of $\si{\micro C}/\si{cm}^2$. Both the size of the squares, $b$, and the distance between them, $d$, are chosen to be much larger than $\beta$, where $\beta$ is the radius characterizing the long-range part of the PSF. These conditions ensure that in the center of the squares, the deposited doses are equal to the uniform incident doses. (b) Bright-field optical micrograph of the device after exposure and development, where the remaining resist thickness determines the color. (c) Dark-field optical micrograph of the same region as in (b). The overlapping subfields from the exposure are visible as a mesh for doses close to the clearing dose. (d)-(e), Cutouts of the central 50$\times\SI{50}{\mu m}^2$ regions of the exposed squares. (f) Extraction of the clearing dose as the minimum dose where the intensities from (d) and (e) are independent of dose, which is consistent across both the bright- and dark-field data sets.
    }
\end{figure*}

Turning to the experiment, we first expose a set of large squares with different uniform doses as illustrated in Fig.~\ref{fig:ClearingDose}(a). 
Both the size of the squares, $b$, and the distance between them, $d$, are chosen to be much larger than the radius characterizing the long-range part of the PSF, $\beta$. This ensures that the measured clearing dose is pattern-independent and that the deposited dose in the center is equal to the uniform incident dose. For an acceleration voltage of $V=\SI{150}{kV}$ and a silicon substrate, then $\beta\sim\SI{60}{\micro m}$, so we choose $b=\SI{250}{\micro m}$.
Figure~\ref{fig:ClearingDose}(b-c) shows bright- and dark-field optical-microscopy (OM) images of the exposed and developed pattern with insets for $D_{\mathrm{in}}=\SI{225}{\mu C/cm^2}$.
The median of the normalized pixel intensity $I_{\mathrm{p}}$ is computed from the central cutouts of Fig.~\ref{fig:ClearingDose}(d-e). 
This is used to determine the clearing dose, $D_0=225\pm\SI{5}{\mu C/cm^2}$, as the dose with no remaining resist, \textit{i.e.}, we observe no change in pixel intensity when increasing the dose further, based on an empirical condition for the gradient, $\nabla I_{\mathrm{p}} \leq 0.05$, being fulfilled for $D_{\mathrm{in}}\geq D_0$, see Fig.~\ref{fig:ClearingDose}(f).

\subsection{Measurement of the process point-spread function from point exposures}

\begin{figure*}[h]
    \centering
    \includegraphics[width=0.85\linewidth]{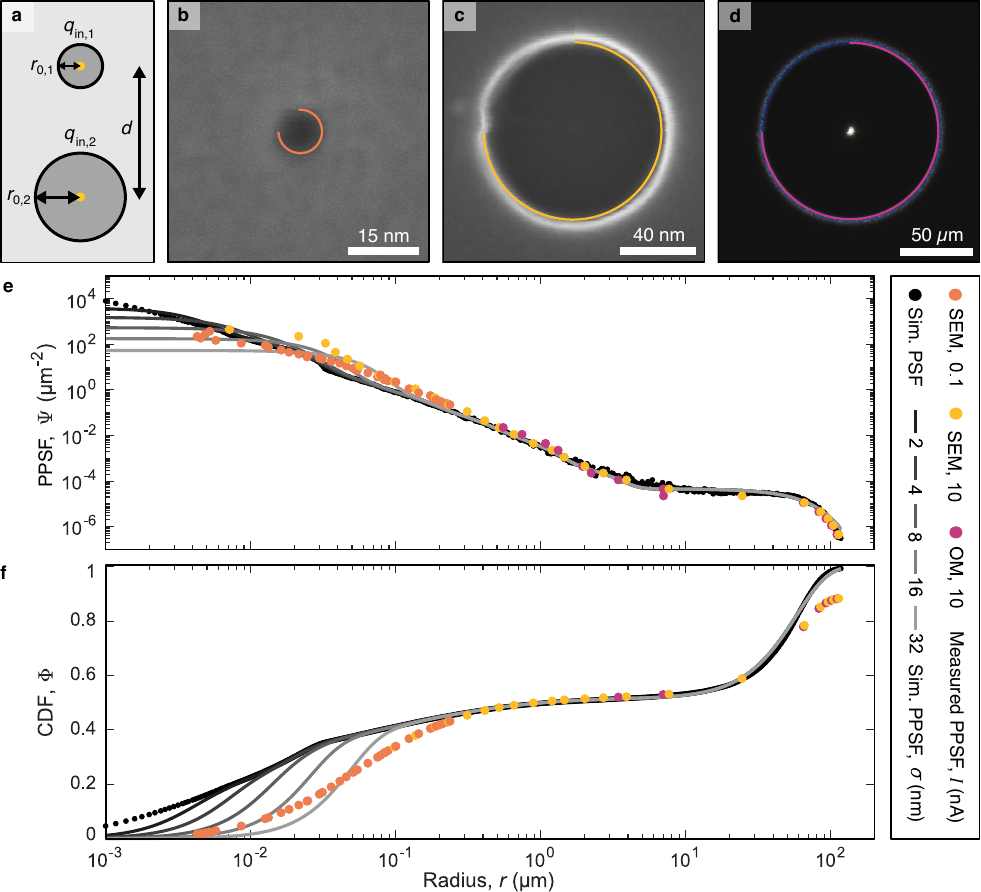}
    \caption{\label{fig:PSFexperiment}
    Measurement of the process point-spread function (PPSF). We consider the same parameters as in Fig.~\ref{fig:ClearingDose}, \textit{i.e.}, \SI{150}{kV} exposures on a stack comprising \SI{180}{nm} CSAR on silicon. (a) Illustration of two isolated points exposed with shot charges, $q_{\mathrm{in},1}<q_{\mathrm{in},2}$, such that the resist is fully developed out to radii of $r_{0,1}<r_{0,2}$. To avoid the two areas affecting each other through the proximity effect, they are separated by a distance $d\gg r_{0,2}$. (b)-(c) Scanning electron microscopy (SEM) and (d) optical microscopy (OM) images of the developed patterns. The exposed holes are fitted to obtain the outline (colored 3/4 circles) and radii. 
    (e) The deposited charge as a function of the measured radius, \textit{i.e.}, the PPSF obtained by varying the deposited charge (colored dots). Two different currents, $I$, are used to reduce the writing time for very large radii. Except for a few points at shorter radii, the analysis by OM and SEM give very similar results, which in turn agree very well with the the PSF extracted from a Monte Carlo simulation (black dots). 
    The deviations for radii below \SI{50}{nm} could originate from process effects. By convolving the PSF with Gaussians of different standard deviations, $\sigma$, simulated PPSFs that include the beam distribution are obtained (solid lines).
    (f) The cumulative density functions (CDFs) of the PPSFs emphasize the deviations, as well as the varying contributions to the deposited dose across the PPSFs.
    }
\end{figure*}

Having determined $D_0$, we now consider the measurement of the PPSF according to Eq.~(\ref{eq:psf_measurement_theory}). We therefore expose sequences of points as illustrated schematically in Fig.~\ref{fig:PSFexperiment}(a), which shows two point exposures of charge $q_{\mathrm{in},1}<q_{\mathrm{in},2}$, separated by a distance $d$ large enough to consider each exposure isolated, such that development removes resist out to radii $r_{0,1}<r_{0,2}$. We then use scanning electron microscopy (SEM) and OM images of the developed holes to measure the radii $r_{0}$ as shown in Fig.~\ref{fig:PSFexperiment}(b-d). This method benefits from its conceptual simplicity but the resolution of SEM is insufficient to accurately measure holes resulting from impinging shot charges below $q_{\mathrm{in}}=\SI{6}{fC}$. For high doses, a central pillar becomes cross-linked as can be seen in Fig.~\ref{fig:PSFexperiment}(d) and may even topple over, which is apparent from a careful inspection of Fig.~\ref{fig:PSFexperiment}(c). To extract the radii of the developed holes, we use image analysis as explained in Appendix~\ref{app:image_analysis}.

The measured PPSF is shown in Fig.~\ref{fig:PSFexperiment}(e). In order to reduce the writing time for the longer exposures, two different currents, $I=\SI{0.1}{nA}$ and $I=\SI{10}{nA}$, are used for the experiment. We observe that the measured PPSF at $I=\SI{10}{nA}$ is higher than at $I=\SI{0.1}{nA}$ in the range of \SI{5}{nm}-\SI{50}{nm}. We note that this cannot be explained solely by differences in beam waist resulting from the different apertures used for the two currents, because such broadening should be included as convolutions that would preserve the normalization. Instead, we observe a systematically higher PPSF for $I=\SI{10}{nA}$ such that its integral is not preserved. This could indicate that the development rate depends weakly on the exposure current, which is not included in the theoretical framework outlined above, and could be attributed to radiation damage or similar effects. This shows that, at least for CSAR resist, one must be careful when switching between currents. 
In practice, lower currents are used for small features because of intrinsically smaller spot sizes and minimum-dwell-time limits, so we consider the PPSF at $I=\SI{0.1}{nA}$ more representative at short ranges. Additional details on the data analysis is included in Appendix~\ref{app:data_extra_info}.

Figure~\ref{fig:PSFexperiment}(e) shows also the PSF obtained from Monte Carlo simulations. We observe excellent agreement except at very short radii, given that the data set extends over so many orders of magnitude. For the same reason, the apparent agreement may be misleading, as a few experimental points actually deviate from the simulation by more than an order of magnitude.
In order to investigate whether the finite beam waist is responsible for the deviations between the simulated PSF and the measured PPSF at small radii, we convolve the PSF with Gaussians of different widths as shown in Fig.~\ref{fig:PSFexperiment}(e). While this provides a slightly better agreement with the PPSF, the lack of data points below \SI{4}{nm} make this comparison inconclusive. 

The several orders of magnitude covered by the PPSF hide its nuances, some of which become clearer by also considering its cumulative density function (CDF),
\begin{equation}\label{eq:cdf}
    \Phi(r) = 2\pi \int_0^{r} \Psi(\tilde{r})\, \tilde{r} \,d\tilde{r},
\end{equation}
which represents the fraction of the total dose deposited out to a given range. Importantly, simulations or models of the PPSF also need to accurately represent the CDF, which is complicated by the fact that seemingly minor changes in the PPSF, due to its scale, can significantly impact the CDF.
The CDF estimated by trapezoidal numerical integration of the data in Fig.~\ref{fig:PSFexperiment}(e) is shown in Fig.~\ref{fig:PSFexperiment}(f). Notably, the integral of the measured PPSF, \textit{i.e.}, the maximum value of the CDF, is $\mathrm{max}(\Phi)=0.88$. Shot noise adds uncertainty to $q_{\mathrm{in}}$, but this noise averages out to less than 1\% for all doses measured in our experiment. Nevertheless, multiple other factors could explain the deviation from the ideal value of unity: One is the above-mentioned current-dependency. A second is numerical errors associated with integrating a function over so many orders of magnitude. A third is errors in the measurement of $r_{0}$, which may be biased by, \textit{e.g.}, sidewall inclinations, line-edge roughness, and SEM artefacts. This is particularly pronounced for very small and very large radii. Finally, hydrodynamic effects may impact the development for the smallest radii as diffusion in and out of the tiny holes is slowed down, or impact the measurement of $D_0$ due to forces at the resist-substrate interface. In the rest of this paper, we normalize the measured CDFs as $\Phi/\mathrm{max}(\Phi)$ to evenly distribute errors from any of these effects.

\subsection{Models of the process point-spread function at high acceleration voltages}

\begin{figure*}[h]
    \centering
    \includegraphics[width=0.84\linewidth]{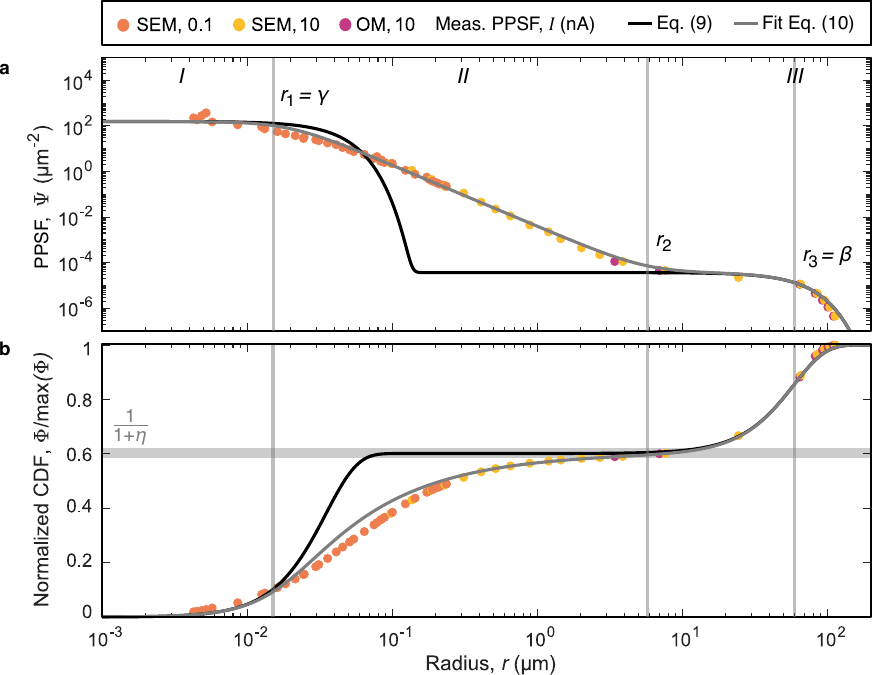}
    \caption{\label{fig:PSFmodels}
    Comparison of theoretical models of the process point-spread function (PPSF).
    We consider the same parameters as in Fig.~\ref{fig:ClearingDose}, \textit{i.e.}, \SI{150}{kV} exposures on a stack comprising 180 nm CSAR on silicon.
    (a) The measured PPSF (orange, yellow, and purple dots) is fitted with a power-law plus Gaussian (PLG) model (grey line) with excellent agreement. The commonly employed double-Gaussian (2G) model (black line) deviates from measurements by up to four orders of magnitude.
    The three regions \textit{I}-\textit{III} introduced in Fig.~\ref{fig:PSFintro}(d) are well-defined in the PLG model: The radius at half maximum, $\gamma=r_1$, the radius at which the two model terms are equal, $r_2$, and the characteristic radius of the long-range Gaussian term, $\beta=r_3$.
    (b) The corresponding cumulative density functions (CDFs) of the PLG and 2G models coincide for $r>r_2$.}
\end{figure*}

For a quantitative description of the PPSF, the model most commonly used is the 2G model,
\begin{equation}
    \Psi_\text{2G}(r) = \frac{1}{\pi(1+\eta)} \left[ \frac{1}{\alpha^2}\text{e}^{-\frac{r^2}{\alpha^2}} +
    \frac{\eta}{\beta^2}\text{e}^{-\frac{r^2}{\beta^2}} \right],
\end{equation}
where the standard deviations $\alpha/\sqrt{2}$ and $\beta/\sqrt{2}$ respectively represent the dose deposited by the forward-scattered electrons and the electrons back-scattered from the substrate into the resist, with $\eta$ describing their relative contribution. 
The advantages of the 2G model include simplicity and numerical speed. It is also possible to extract reasonable estimates of $\eta$ and $\beta$ for PEC by fitting the 2G model to the long-range PPSF or CDF while setting $\alpha=0$. Figure~\ref{fig:PSFmodels} shows the same data as in Fig.~\ref{fig:PSFexperiment} compared to different models of the PPSF.
The 2G model agrees well with the PPSF for large radii but fails at shorter radii, in particular in the intermediate range where it deviates by up to four orders of magnitude.

On this background, we propose replacing the forward Gaussian with a power law, resulting in a power-law plus Gaussian (PLG),
\begin{equation}
    \label{eq:power_law_model}
    \Psi_{\mathrm{PLG}}(r) = \frac{1}{\pi(1+\eta)} \left[ \frac{\nu^{-2}}{\left(\frac{r}{\gamma}\right)^\sigma + \text{e}^{ \frac{r^2 }{ \left(\beta/4 \right)^2}}} + 
    \frac{\eta}{\beta^2}\text{e}^{-\frac{r^2}{\beta^2}}  \right],
\end{equation}
where $\beta/\sqrt{2}$ remains the standard deviation of the long-range scattering, $\eta$ is now the ratio of long- to short- and intermediate-range scattering (both can be inserted directly into a 2G model), $\gamma$ is the radius at half maximum of the forward distribution and marks the beginning of the power-law scaling, $\sigma$ is the slope of the power law and $\nu$ is the normalization constant of the power-law term. 
Our model builds on and extends previous efforts to model the PPSF with a power-law~\cite{albrechtsen_power-law_2019,mao_proximity_2025}, including an empirical Gaussian cut-off term in the denominator of the power-law term to ensure both that the long-range Gaussian dominates at $r\gg\beta$ and to avoid a singularity at $r=\SI{0}{nm}$.

The parameters of the PLG model define the regions \textit{I}-\textit{III}: The short-range limit is $r_1=\gamma$, and the radius at which the two model terms are equal, \textit{i.e.}, the intermediate- to long-range boundary is $r_2$. The regions are indicated in Fig.~\ref{fig:PSFmodels} together with the characteristic radius of the long-range distribution, $r_3=\beta$, and a fit of the PLG model to the measured PPSF, which show excellent agreement. Minor systematic deviations of the fit are present in the short-to-intermediate range and beyond $\beta$, but such deviations are to be expected, since the PLG model remains empirical and cannot fully capture the underlying physics of electron scattering. In any case, the agreement over many orders of magnitude in both radii and PPSF is remarkable.
For comparison, the 2G model in Fig.~\ref{fig:PSFmodels} is chosen to have the same values of $\eta$ and $\beta$ as the PLG model fit, as well as a value for $\alpha$ that replicates the short-range behavior. We refer to Appendix~\ref{app:plg_fits} for further details on the fitting procedure.

Importantly, the PLG model also accurately captures $\eta$ in the CDF. Here the maximum value of the power-law term, $\Phi=(1+\eta)^{-1}$, clearly distinguishes the short- to intermediate-range dose contribution from that of the long range. This means that we can consider regions \textit{I}-\textit{II} and \textit{III} separately for PEC: When patterns are fractured into polygons of sizes larger than $r_2$, the proximity effect from one polygon to another is solely governed by the long-range Gaussian, and the proximity effect of the polygons onto themselves, \textit{i.e.}, the so-called self-dose, is solely governed by the short-to-intermediate ranges and reduces to $(1+\eta)^{-1}$. This is a too coarse fracturing for high-resolution lithography, but it can be applied away from polygon edges. Conversely, the long-range Gaussian does not contribute to the self-dose for polygons smaller than $r_2$.
For practical applications, the PLG model can give an accurate estimate of the $\eta$ parameter, which is essential to many PEC algorithms, but its main purpose is as a simple model to allow for detailed comparisons between different material stacks, acceleration voltages, and process parameters.

\subsection{Measurements of the 150 kV process point-spread functions for longer development, aluminum decharging layer and indium phosphide substrate}

\begin{figure*}[h]
    \centering
    \includegraphics[width=0.87\linewidth]{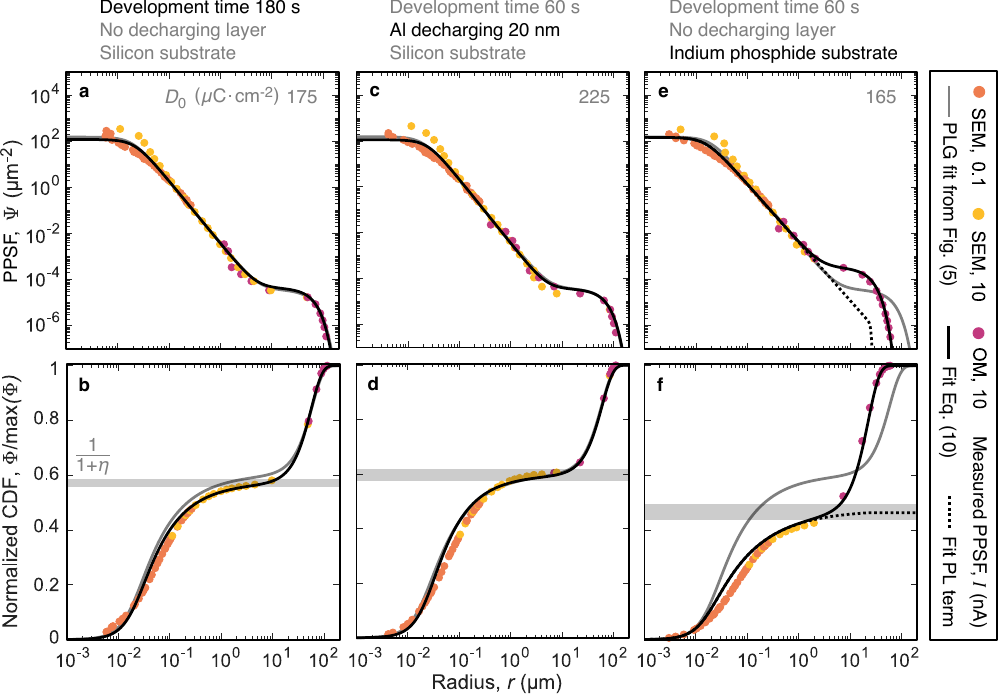}
    \caption{\label{fig:PSFcomparison}
    Process point-spread functions (PPSFs) and the corresponding cumulative density functions (CDFs) for \SI{150}{kV} exposures on \SI{180}{nm} CSAR with varying material stacks and processing parameters. Measured PPSFs (colored dots) are well reproduced by fits (black lines) with a power-law plus Gaussian (PLG). The clearing dose, $D_0$ extracted from a corresponding dose test (not shown) is indicated for each experiment. The PLG fit to the results presented in Figs.~\ref{fig:ClearingDose}-\ref{fig:PSFmodels} is shown for comparison (grey lines). (a)-(b) An increased development time lowers the clearing dose, but affects the PPSF minimally. (c)-(d) A \SI{20}{nm} aluminum decharging layer on top of the resist during electron-beam exposure results in a negligible change. (e)-(f) An indium phosphide substrate instead of silicon significantly decreases the clearing dose and changes the PPSF: The long-range contribution falls off at shorter radii, but contributes relatively more to the deposited dose as seen from the CDF, which is attributed to increased electron scattering in the denser substrate.
    }
\end{figure*}

In order to compare the effects of different material systems and process parameters, we measure the PPSF in three additional experiments, each one modifying one parameter:
\begin{enumerate}[noitemsep]
    \item Tripling the development time, which should reduce the clearing dose, $D_0$, but retain the same deposited dose distribution, $D(r)$, and thus the same PSF.
    \item Adding a \SI{20}{nm} aluminum layer on top of the resist to reduce any effects from deposited charges during exposure.
    \item Exchanging the substrate to InP, which should exhibit increased scattering in the substrate due to its higher density.
\end{enumerate}
The measured PPSFs, their corresponding CDFs, and the fits of the PLG model to these experiments are shown in Fig.~\ref{fig:PSFcomparison}, together with the fit of the PLG model to the experiment shown in Figs.~\ref{fig:ClearingDose}-\ref{fig:PSFmodels}.
The fit parameters for each of our experiments are summarized in table~\ref{tab:psf_alternate}.

\begin{table*}[h]
\centering
\caption{\label{tab:psf_alternate} Measured clearing doses, $D_0$, integrals of the process point spread function (PPSF), max($\Phi$), and fit parameters of the power-law plus Gaussian (PLG) model, $\beta$, $\eta$, $\gamma$, $\sigma$, and $\nu$.}
\begin{tabular}{l c c c c c c c}
\hline
~ & $D_0$ & max($\Phi$) & $\beta$ & $\eta$ & $\gamma$ & $\sigma$ & $\nu$\textsuperscript{a}\\
~ & \si{\micro\coulomb\per\centi\meter\squared} & ~ & \si{\micro\meter} & ~ & \si{nm} & ~ & \si{nm}\\
\hline
Si    & 225(5) & 0.88 & 59(4) & 0.67(5) & 15(2) & 2.69(7) & 24.9\\
Si L\textsuperscript{b} & 175(5) & 0.89 & 56(3) & 0.75(4) & 17(2) & 2.71(7) & 27.3\\
Si Al & 225(5) & 0.84 & 60(5) & 0.68(6) & 19(3) & 2.79(10) & 28.9\\
InP   & 165(5) & 0.93 & 23(1) & 1.16(13) & 12(2) & 2.52(9) & 22.2\\
\hline
\end{tabular}
\\\textsuperscript{a} $\nu$ values required for normalization with $\gamma$ and $\sigma$ best estimates.\\
\textsuperscript{b} Longer development time of \SI{180}{s} instead of \SI{60}{s}.
\end{table*}

In the experiment with a longer development time, more resist is dissolved, so each hole of a given shot dose, $q_{\mathrm{in}}$, is developed out to a larger radius, $r_{0}$. However, this is offset by a reduction in the clearing dose, $D_0$, such that the PPSF remains approximately the same, see Fig.~\ref{fig:PSFcomparison}(a)-(b).

Adding the decharging layer results in a slight broadening of the short-range PPSF, as shown in Fig.~\ref{fig:PSFcomparison}(c)-(d). The difference between currents remains. On the one hand, this indicates that this difference does not originate from charging effects and, on the other hand, that the effect of the thin decharging layer is negligible for exposure of thin resists at high acceleration voltages.

The long-range part of the PPSF is significantly different for the InP substrate as shown in Fig.~\ref{fig:PSFcomparison}(e)-(f), whereas the short- to intermediate-range distributions remain largely invariant between substrates. 
The decrease in $\beta$ highlights that, without the power-law cutoff term in the PLG model, the power-law term would dominate beyond $\beta$ and result in a poor fit. 
Additionally, $\eta$ increases significantly, which is evident from the CDF, and also leads to a decrease in $D_0$. Essentially, the long-range scattering is stronger but shorter due to the increased electron scattering on the heavier elements in the InP substrate.

\subsection{The effect of acceleration voltage on the process point-spread function}

\begin{figure*}
    \centering
    \includegraphics[width=0.88\linewidth]{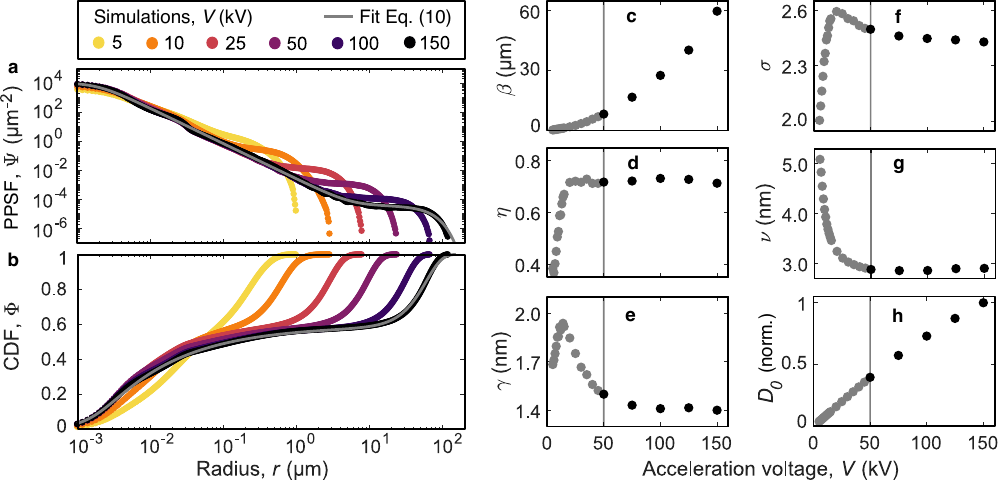}
    \caption{\label{fig:PSFvoltages}
    The dependence of the process point-spread function (PPSF) on acceleration voltage, $V$, in the range $V=5$-\SI{150}{kV} for \SI{180}{nm} CSAR on silicon.
    (a) Estimate of the PPSF for different voltages (colors). These are calculated by convolving the point-spread functions obtained from Monte Carlo simulations with a Gaussian of full width at half maximum \SI{4}{nm} approximating the finite beam waist.
    The fit to the power-law plus Gaussian (PLG) model for $V=\SI{150}{kV}$ is also shown (grey line).
    (b) The cumulative density functions (CDFs) corresponding to the PPSFs.
    (c)-(g) The fitting parameters of the PLG model as a function of acceleration voltage. (h) The clearing dose calculated from the normalization of each simulation, normalized to $D_0(V_0)$ where $V_0=\SI{150}{kV}$, \textit{cf}.\ Eq.~\ref{eq:ClearingDose_vs_voltage}. We observe that $\beta$ and $D_0$ increase with $V$, while the remaining parameters, $\eta$, $\gamma$, $\sigma$, and $\nu$ remain approximately constant for $V>\SI{50}{V}$ (black dots) and vary significantly for $V<\SI{50}{V}$ (gray dots), which we attribute to three-dimensional effects as discussed in the main text.
    }
\end{figure*}

We further explore the applicability of the PLG model for various acceleration voltages, $V$.
We simulate the PSFs for \SI{180}{nm} CSAR on silicon with $V$=[5-150]~\si{kV} and convolve them with a Gaussian with a full width at half maximum of \SI{4}{nm} in order to estimate the PPSFs. The result is shown in Fig.~\ref{fig:PSFvoltages}(a) with the corresponding CDFs shown in Fig.~\ref{fig:PSFvoltages}(b). 
We observe that the short-range contribution is much less dependent on $V$ than the long-range part. A fit of the PLG model for \SI{150}{kV} is included as an example, and we refer to Appendix~\ref{app:plg_fits} for fits at other voltages and for comparisons with the popular resists PMMA and HSQ. The fits capture the PPSFs accurately across all voltages and resists, and we extract the parameters $\beta$, $\eta$, $\gamma$, $\sigma$, and $\nu$ as a function of $V$, as shown in Fig.~\ref{fig:PSFvoltages}(c)-(g).

We note two distinct regimes, wherein the parameters $\eta$, $\gamma$, $\sigma$, and $\nu$ remain approximately constant for $V>\SI{50}{kV}$ but vary significantly for $V<\SI{50}{kV}$. Conversely, $\beta$ follows an approximately quadratic dependence across all acceleration voltages. At the lower voltages, $\beta$ reduces from a few microns to a few hundred nanometers, such that the PPSFs likely vary significantly within the depth of the resist. Thus, we attribute the behavior at $V<\SI{50}{kV}$ to the fact that we average out the depth dependence of the PSFs in the output of our simulations. In contrast, when $V>\SI{50}{kV}$ the scattering length is high compared to the resist thickness and the only variation with $V$ is the increase in $\beta$, which simply shows that the same fraction of long-range dose is distributed across a larger area. The consequence of this is that PEC can be simpler for large $V$ because the long-range scattering is approximately constant for patterns much smaller than $\beta$. However, the larger range also implies that PEC must correct for proximity doses over a much larger area.
The remaining PLG model parameters stay constant at high $V$ and highlight that the short- to intermediate-range proximity effects remain relevant for nanostructure fabrication. This is problematic for PEC based on the 2G model, as it assumes the intermediate range is negligible.

Lastly, it should be noted that increasing $V$ also reduces the total energy deposited by scattering, $E_{\mathrm{tot}}(V)=2\pi \int E(r,V) \, r\,dr$, which is the normalization constant of the PSF simulations, and inversely related to the clearing dose as
\begin{equation}\label{eq:ClearingDose_vs_voltage}
    \frac{D_0(V)}{D_0(V_0)} = \frac{E_{\mathrm{tot}}(V_0)}{E_{\mathrm{tot}}(V)},
\end{equation}
with $V_0$ some fixed voltage. Hence, simulations of the electron scattering give estimates of $D_0(V)$, as shown in Fig.~\ref{fig:PSFvoltages}(h) with $V_0=\SI{150}{kV}$. The nearly linear dependence of $D_0(V)$ means that high acceleration voltages enable finer dose control at the cost of speed, which can be necessary in order to achieve increasingly smaller feature sizes.

\subsection{The effect of the short- to intermediate range point-spread function on lithography}

\begin{figure*}[h!]
    \centering
    \includegraphics[width=0.9\linewidth]{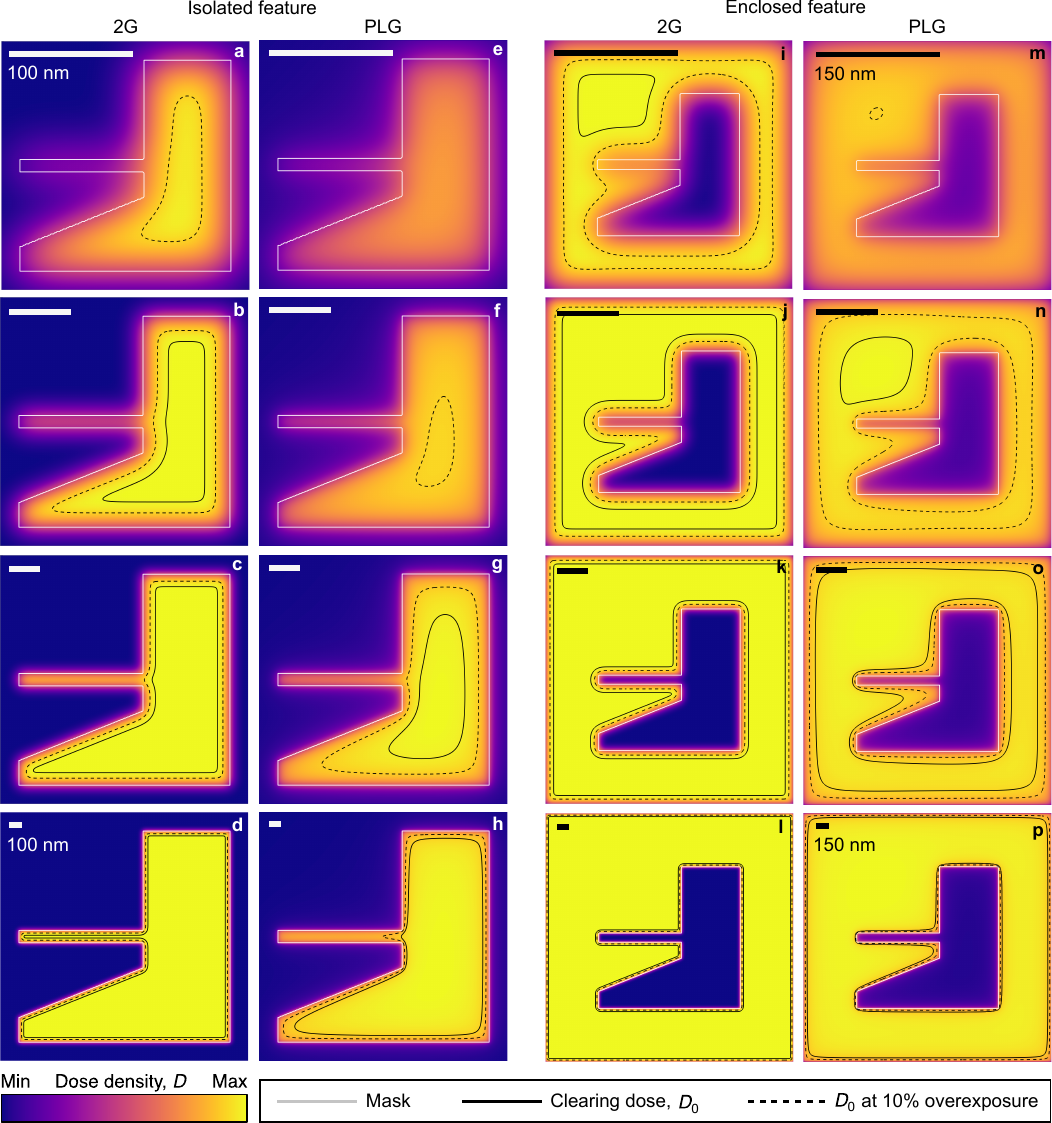}
    \caption{\label{fig:PSFconvolutions}
    Impact of the point-spread function on electron-beam lithography at different length scales.
    (a-d) Deposited dose density, $D(x,y)$, for increasing feature sizes when the mask of an isolated feature (exposure within white outline) is convoluted with the double-Gaussian (2G) process point-spread function (PPSF) from Fig.~\ref{fig:PSFmodels}. The threshold at which resist is fully cleared after development, $D_0=D(x,y)$, is indicated when the incident dose is $D_\mathrm{in}=D_0(1+\eta)$ (full black line) and $D_\mathrm{in}=1.1D_0(1+\eta)$ (dashed black line), respectively. The factor $(1+\eta)$ compensates that the long-range scattering deposits dose almost entirely outside of the visible areas.
    (e-h) The same mask convoluted with the power-law plus Gaussian (PLG) PPSF from Fig.~\ref{fig:PSFmodels}. Less dose is deposited in the range tens-of-nanometers, but more in the range hundreds-of-nanometers, leading to a more pronounced proximity effect. As feature sizes increase, both the disparity between PPSFs and distortions due to the proximity effect become less visible. 
    (i-l) Enclosed versions of the mask (exposure from white line to border) convoluted with the same 2G PPSF, 
    (m-p) and PLG PPSF. Both isolated and enclosed features experience similar distortions, but the contrast in $D(x,y)$ is worse for enclosed features, making them particularly difficult to resolve in lithography.
    }
\end{figure*}

Finally, we consider how our findings impact EBL at different scales. The trend towards narrower beam widths and higher $V$ results in substantial deviations from the 2G-model at $\SI{150}{kV}$ for feature sizes on the order of tens- to hundreds of nanometers, i.e., all nanoscale structures. The impact of this deviation for patterns with feature sizes \SI{10}{nm}-\SI{1}{\micro m} is visualized in Fig.~\ref{fig:PSFconvolutions}, which shows $D(x,y)$ of masks convolved with the 2G or PLG PPSFs from Fig.~\ref{fig:PSFmodels}. The exposed areas are much smaller than $r_2$, so the dose deposited from long-range scattering is almost entirely outside of the visible areas of the plots, giving an approximately constant reduction of the deposited dose. This means that only the short- to intermediate-range of the PPSFs give rise to spatial variations in the deposited dose. In all cases, the areas that would be cleared of resist after development (black lines) do not match the masks (white lines), because the PPSFs distribute dose away from the points of exposure --- the fundamental problem that PEC addresses and corrects for.

The relative difference between mask and result is larger for smaller patterns, since the length scale of the deviation is set by the PPSF, and it is therefore independent of the feature size and remains on the order of \SI{100}{nm} or below: This corresponds to the length scale at which the PPSFs change significantly and the deviation between PPSFs for 2G- and PLG-models is largest at \SI{150}{kV}. The deviation between the mask and the resulting pattern is significantly larger for the PLG model, and, since the PLG model is a much better model of the PPSFs than the 2G model, this shows that the 2G-model significantly underestimates the magnitude of the proximity effect. This, in turn, implies that PEC based on the 2G-model would not only underestimate the proximity effect, but it would also lead to distorted patterns with poor agreement with the mask.

\section*{Conclusions}

We have measured the PPSF that characterizes the proximity effect in EBL. The PPSF, obtained from exposures of single shots and normalized by the pattern-independent clearing dose, shows excellent agreement with Monte Carlo simulations of the PPSF and CDF, although we observe deviations in the regions limited by metrology or the number of data points. 

When applying the common 2G model of the PPSF, it deviates from experiments by more than four orders of magnitude because it disregards the intermediate range. We therefore propose a PLG model that shows excellent agreement with all the experiments and Monte Carlo simulations of the PPSFs, independent of material stacks, development time, and acceleration voltages in the range of $5-\SI{150}{kV}$. This indicates that the PLG model can capture the key features of the PPSFs regardless of all microscopic details, while simultaneously offering the advantages of a simple model with interpretable parameters. This allows comparisons between exposures, processes, and material platforms in order to obtain insights into the electron scattering and PEC.

We find that it is only meaningful to assign $\eta$ to distinctly separate the long-range dose contribution of the PPSF from the short- to intermediate-range contributions when the electron acceleration voltage is high compared to the resist thickness and electron scattering in the substrate. For a stack consisting of \SI{180}{nm} resist on silicon, this separation is excellent for $V\geq\SI{50}{V}$. 
The boundary is captured by the radius at which the two terms of the PLG model are equal, i.e, when the long-range approximation deviates from the PPSF by a factor of two, which far exceeds the forward-scattering approximation of the 2G model. The complicated behavior of the PPSF at low voltages, which is likely due to three-dimensional effects, calls for further simulations and experiments. The clearing dose scales nearly linearly with voltage, enabling finer dose control at the cost of speed.

In the high-voltage regime, the short- to intermediate-range PPSF is accurately described by a power law and it is largely unaffected by voltage, material stack, and development time. However, $\eta$ remains constant for a given substrate while $\beta$ increases with an approximately square dependence, implying that the relative long-range dose contribution remains the same, only distributed across longer distances. This potentially simplifies PEC because the long-range dose is approximately constant for distances $\ll\beta$, although larger areas need to be accounted for.
However, at the short- to intermediate range, the 2G model is increasingly inaccurate at higher voltages. Thus, the tendency towards the intuitively and mathematically simple 2G model in PEC has become a limiting factor, specifically in high-resolution EBL, where PEC based on accurate PPSFs from numerical simulations, many-Gaussian approximations, or PLG-like models could improve device fabrication.

\ack{
S.S. acknowledges Bingrui Lu for valuable discussions.
}

\funding{
We gratefully acknowledge financial support from the Villum Foundation Young Investigator Program (Grant No.\ 13170), Innovation Fund Denmark (Grants No.\ 0175-00022 -- NEXUS), the European Research Council (Grant No.\ 101045396 -- SPOTLIGHT), the Horizon Europe Research and Innovation Programme (Grant No.\ 101098961 -- NEUROPIC, and the Danish National Research Foundation (Grant No.\ DNRF147 -- NanoPhoton).
}

\section*{Conflicts of interest}
N.B.H., K.S.K., M.A., T.S., and C.A.R. declare no conflict of interests. S.S. declares his partial ownership of Beamfox Technologies ApS.

\roles{
Nikolaj Balslev Hougs\orcid{0009-0000-2625-1201}
\newline
Writing\textemdash original draft, Writing\textemdash review and editing, Visualization, Methodology, Formal analysis, Data curation, Conceptualization.
\vspace{0.15cm}\newline
Kristian Sofus Knudsen
\newline
Writing\textemdash original draft, Writing\textemdash review and editing, Visualization, Methodology, Formal analysis, Data curation, Conceptualization.
\vspace{0.15cm}\newline
Marcus Albrechtsen\orcid{0000-0003-4226-0997}
\newline
Writing\textemdash original draft, Writing\textemdash review and editing, Methodology, Data curation, Conceptualization, Supervision.
\vspace{0.15cm}\newline
Taichi Suhara
\newline
Methodology, Data curation, Conceptualization.
\vspace{0.15cm}\newline
Christian Anker Rosiek\orcid{0000-0003-3362-4752}
\newline
Writing\textemdash review and editing, Conceptualization.
\vspace{0.15cm}\newline
Søren Stobbe\orcid{0000-0002-0991-041X}
\newline
Writing\textemdash review and editing, Methodology, Conceptualization, Project administration, Funding Acquisition, Supervision.
}

\data{
Data from the article is freely available at Zenodo, https://zenodo.org/records/15691545.
}

\appendix
\section*{Appendices}
\section{Fabrication\label{app:fabrication}}

Samples consisting of \SI{180}{nm} AllResist (Strausberg, Germany) AR-P 6200 (CSAR) on silicon and InP substrates are exposed using a \SI{150}{kV} Elionix (Tokyo, Japan) ELS-F150 electron-beam writer. Each sample contains both point and area exposures: The pattern of point exposures is one array of shot charges $q_{\mathrm{in}}$ ranging from $\SI{1}{fC}$ to $\SI{5}{\micro C}$ at pitch $d=\SI{140}{\micro m}$ and beam current $I=\SI{10}{nA}$, and an additional array with $q_{\mathrm{in}}$ from $\SI{1}{fC}$ to $\SI{10}{pC}$ with $d=\SI{5}{\micro m}$ and $I=\SI{0.1}{nA}$. The pattern of area exposures is an array of squares of side lengths $b=\SI{250}{\micro m}$, receiving a uniform exposure dose $D_{\mathrm{in}}$ in the range $180\si{\micro C \cdot cm^{-2}}$ to $235\si{\micro C \cdot cm^{-2}}$.

The samples are developed in single-puddle Zeon Corporation (Tokyo, Japan) ZED-N50 developer for \SI{60}{s} before the developed radii, $r_{0}$, are measured using scanning electron microscopy (SEM) and optical microscopy (OM). 
One sample with silicon substrate is instead developed for \SI{180}{s}. 
Another sample with silicon substrate has a \SI{20}{nm} aluminium decharging layer deposited on top of the resist before exposure. The layer is removed again after exposure, but prior to development of the resist, by development in single-puddle MicroChemicals GmbH (Ulm, Germany) AZ 726 MIF for \SI{60}{s}.

\section{Obtaining the electron point-spread function from Monte Carlo simulations of electron scattering\label{app:psf_from_monte_carlo}}
For this work we have used a Geant4-based Monte Carlo simulation of electron scattering~\cite{agostinelli_geant4simulation_2003, allison_geant4_2006, allison_recent_2016}.
In such a simulation, we consider a resist of depth $z_0$ on a substrate as in Fig.~1(a) of the main text. When simulating the electron scattering, it is typically the energy density deposited in the resist, $\tilde{E}(r,z)$, with $0<z<z_0$, that is obtained. In general, the electron point-spread function (PSF) is then extracted as 
\begin{equation}
    \tilde{\Psi}(r,z) = \frac{\tilde{E}(r,z)}{2\pi \int \int \tilde{E}(r,z) r \,dr\,dz} = \frac{\tilde{E}(r,z)}{\tilde{E}_{\mathrm{tot}}}.
\end{equation}
In cases of thin resist, the PSF is considered invariant in the vertical direction and approximated as an area density
\begin{equation}
    \Psi(r) \approx \int_0^{z_0}\tilde{\Psi}(r,z)\,dz \approx z_0\tilde{\Psi}(r,z_\mathrm{res}),
\end{equation}
with $z_\mathrm{res}$ an arbitrary depth within the resist. This approximation is included in the output of our simulations, except those shown in Fig.~2(b) and (c) of the main text. 
The same approximation can be similarly applied to $\tilde{E}(r,z)\rightarrow E(r)$ in order to obtain Eq.~(1) of the main text.
Note that this changes the units from $[\tilde{\Psi}]=\si{m}^{-3}$ and $[\tilde{E}]=\si{eV}\cdot\si{m}^{-3}$ to $[\Psi]=\si{m}^{-2}$ and $[E]=\si{eV}\cdot\si{m}^{-2}$.

The beam distribution can be included at the simulation stage, but we recommend simulating the deposited energy and PSF from electron scattering of a $\delta$-function beam, $\tilde{\Psi}_{\mathrm{scat}}(r,z)$, after which an arbitrary beam distribution, $\Psi_{\mathrm{beam}}(x,y)$, can be convoluted to find the process-specific PSF (PPSF)
\begin{equation}
    \Psi(x,y) = z_0\tilde{\Psi}_{\mathrm{scat}}(r,z_\mathrm{res}) * \Psi_{\mathrm{beam}}(x,y).
\end{equation}
Any further fabrication steps, such as development or etching, can then be accounted for by further convolutions with appropriate functions, in order to simulate an arbitrary PPSF.

\section{Image analysis to extract the radii of developed holes\label{app:image_analysis}}

\begin{figure*}
    \centering
    \includegraphics[width=0.9\linewidth]{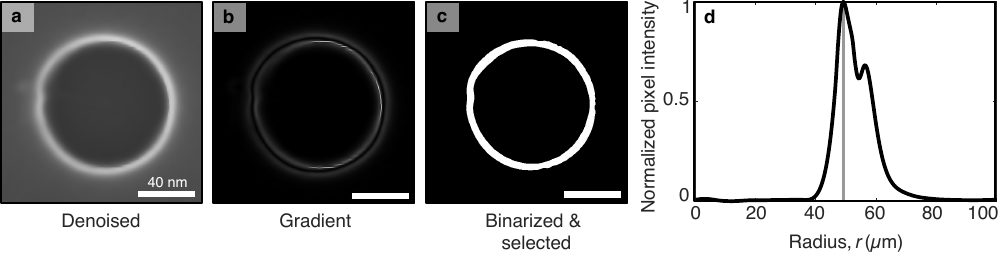}
    \caption{\label{fig:appendixImageAnalysis}Image-analysis process to determine the radii of developed holes. (a) First, the input image is denoised by anisotropic diffusion (100 iterations, local contrast $K = 0.05$, and conduction coefficient weight $\lambda=0.2$)~\cite{perona_scale-space_1990}, (b) then a Sobel filter which produces a discrete approximation of the magnitude of the gradient is applied. (c) Thresholding of the gradient image and removing select pixel groups allows isolation of the circle of interest, whose center is found using a Circular Hough Transform. (d) Lastly, the pixel intensities of the gradient image, integrated wrt.\ radius and normalized, gives peaks corresponding to the inner and outer edge of the bright circle in the input image, where the inner peak is estimated as the resist-substrate interface.}
\end{figure*}

We estimate the radii of holes in the developed resist using an automatic image-analysis process, seeking consistent estimates of radii for the hundreds of images we analyze.
The input data are scanning electron microscopy (SEM) or dark-field optical microscopy (OM) images, such as Fig.~(4)(b)-(d) in the main text. The result is deterministic given the same input image and parameters for the analysis but may be affected by the imaging settings and the slope of the sidewalls. In dark-field OM, topographical variations such as edges or roughness appear bright. In SEM, features appear brighter when the electron beam penetrates material adjacent to an edge, in which case secondary electrons can escape and be detected~\cite{Frase_2009}. Based on this, the inner transition from dark to bright is chosen as the estimate for the developed hole radius, although the true radius may be slightly offset or, in some cases, gradual.
In any case, a rigorous definition of how to extract the hole radius is needed. The following is a summarized description of the automatic image-analysis process:

\begin{enumerate}[noitemsep]
    \item Denoising: Noise in the input image severely hinders the following steps if not removed. Sharp edge features can be preserved while removing high-frequency noise by denoising using anisotropic diffusion~\cite{perona_scale-space_1990}, as shown in Fig.~\ref{fig:appendixImageAnalysis}(a).
    \item Pre-processing: A gradient filter (Sobel) applied to the denoised image results in a gradient image as shown in Fig.~\ref{fig:appendixImageAnalysis}(b). A threshold is applied to binarize the gradient image, then all pixels not connected to the ring-shaped inner edge of the circular hole are removed, as shown in Fig.~\ref{fig:appendixImageAnalysis}(c).
    \item Circle-finding: A Circular Hough Transform, a slow but accurate shape-detection algorithm, is performed on the binary image for various test radii to estimate the circle center and radius. 
    \item Refined estimate: The pixel intensities on the gradient image are integrated radially from the estimated center of the circle, giving graphs as in Fig.~\ref{fig:appendixImageAnalysis}(d) that are resilient against minor errors in the initial center estimates. The final radius is estimated as the peak corresponding to the inner edge of the hole.
\end{enumerate}

\section{Additional details on the data sets and analysis\label{app:data_extra_info}}

We use a low current of $I=\SI{0.1}{nA}$ for the point exposures shown in Figs.~(4)-(6) except for the largest doses where a higher current of $I=\SI{10}{nA}$ was chosen to reduce the write time. 
The two SEM data sets agree well except at very small radii, because the different exposure currents result in physically different hole sizes.
Alternatively, fast and easy measurements devoid of charging artefacts can be obtained from OM, which agrees well with the SEM data for large radii but becomes too inaccurate for smaller radii. We therefore discard the data points for $I=\SI{10}{nA}$ below \SI{50}{nm} and the OM data below $\SI{2}{\mu m}$ when calculating the cumulative density functions (CDFs) and fitting models. 
In Fig.~(4)(f), the OM data point at $\SI{7}{\mu m}$ is inconsistent with the corresponding SEM data point at $\SI{25}{\mu m}$ and is likewise discarded. The discrepancy stems from the different contrast mechanisms in the two imaging methods in the case of a hole with a particularly low and rough sidewall slope.

\section{Power-law plus Gaussian model fits for different acceleration voltages and resists\label{app:plg_fits}}

We first note that power laws are generally difficult to fit well~\cite{clauset_power-law_2009}.
Since the PPSF spans several orders of magnitude from nanometer to micrometer scale, one issue is that simple least-squares fitting is biased towards data points at lower radii, as these have higher values and are more dense, thus producing far larger errors.
The issue stemming from the significant range of the vertical axis can be avoided by fitting the CDF, but this is troublesome without an analytical CDF, as is the case with the PLG model.
Instead, we fit to the PPSF in log-log space to minimize the relative residuals
\begin{equation}
    R = \sum_j \hat{r}_j^2\Delta\hat{r}_j^2[\hat{\Psi}_j(\hat{r}_j)-\hat{\Psi}_{\mathrm{PLG}}(\hat{r}_j)]^2,
\end{equation}
\begin{equation}
    \Delta\hat{r}_j = (\hat{r}_{j+1}-\hat{r}_{j-1})/2,
\end{equation}\\
where $\hat{r}$ indicates $\mathrm{log}_{10}(r)$ and the weights $\Delta\hat{r}_j$ approximate areas of rings with radii $\hat{r}_{j}$ and widths $\Delta\hat{r}_{j}$, as a log-scale approach to the weights used by Rishton and Kern~\cite{rishton_point_1987}.
However, the power-law term of the PLG model is only normalized by selecting a value for $\nu$, so directly fitting the parameters can give incorrect $\nu$ and $\eta$.
For this reason, we first obtain an initial fit that is close to, but not normalized. Then we scale $\nu$ to obtain a normalized but incomplete fit. Lastly, we fit to the CDF to obtain $\eta$, by keeping the remaining parameters from the incomplete fit constant.

\begin{figure*}[h]
    \centering
    \includegraphics[width=0.92\linewidth]{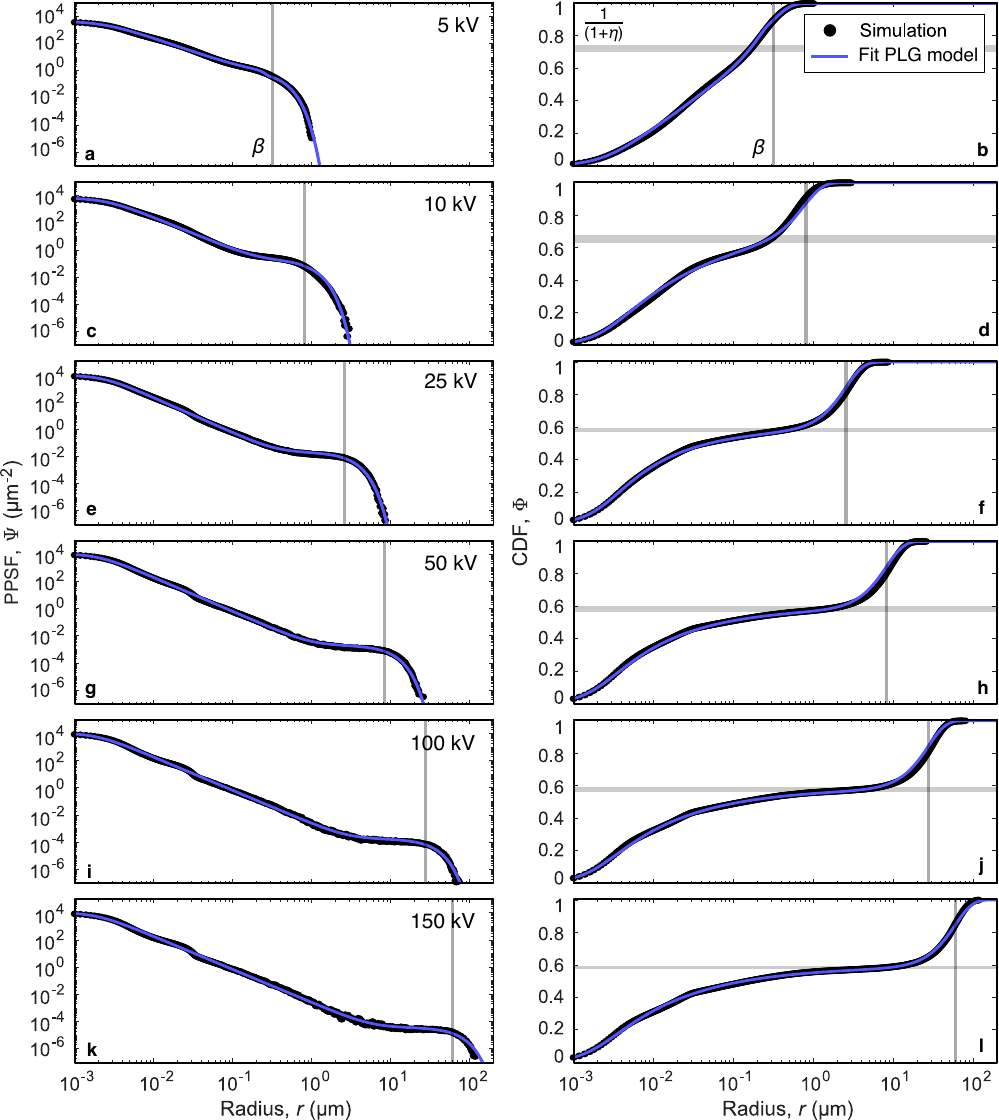}
    \caption{\label{fig:appendixPSFfits}
    Fits to Monte-Carlo simulations of the PPSF with the power-law plus Gaussian (PLG) model. (left) Simulated electron point-spread functions in 180~\si{\nano\meter} CSAR on silicon with increasing electron acceleration voltage, $V$. (right) Corresponding cumulative density functions. The PLG model provides excellent fits in all cases. The vertical gray lines indicate $\beta$ and the horizontal lines indicate the fraction of dose deposited at short to intermediate range, $1/(1+\eta)$. The line thicknesses indicate one standard deviation of the uncertainty in the estimated values. At low $V$, there is a less distinct separation between power-law and Gaussian behavior. As $V$ increases, the model parameter $\beta$ grows and $\eta$ stabilizes.}
\end{figure*}

Figure~\ref{fig:appendixPSFfits} shows examples of such PLG model fits to simulated PPSFs and corresponding CDFs for acceleration voltages $V=5$-\SI{150}{kV}. The fit parameters are included in Table~\ref{tab:sim_fit_parameters}, along with the normalization constant of the simulations. Notably, the two model terms are poorly distinguishable at $5$-\SI{10}{kV}, making estimates for $\eta$ in this range less interpretable as the ratio between a distinct long-range contribution and other dose contributions.

\begin{figure*}[h!]
    \centering
    \includegraphics[width=\linewidth]{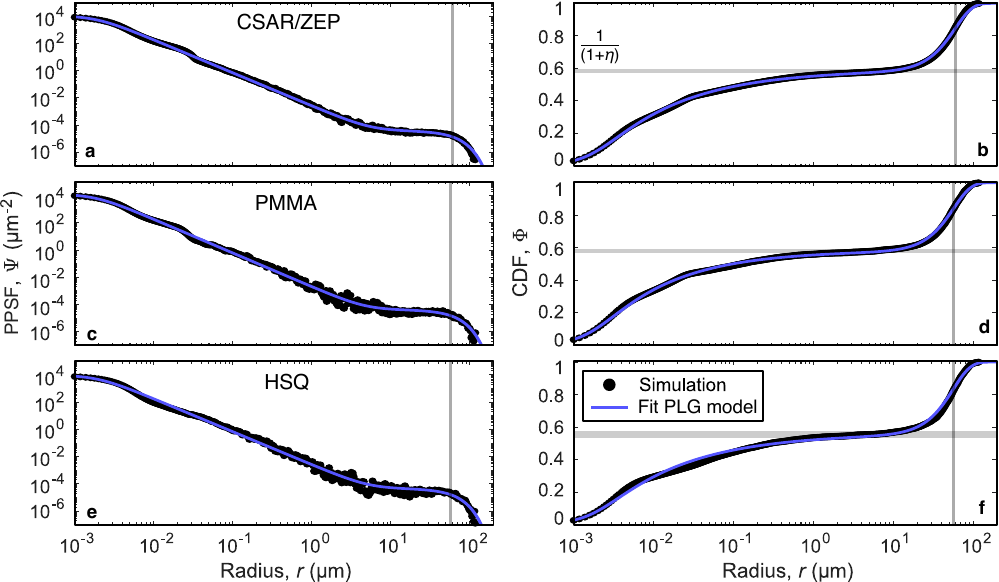}
    \caption{\label{fig:appendixPSFresists}
    Fits to Monte-Carlo simulations of the PPSF with the power-law plus Gaussian (PLG) model for different resists. Left column: Simulated electron point-spread functions in 180~\si{\nano\meter} resist on silicon with \SI{150}{kV} electron acceleration voltage. Right column: The corresponding cumulative density functions. The PLG model provides excellent fits in all cases. The vertical gray lines indicate $\beta$ and the horizontal lines indicate the fraction of dose deposited at short to intermediate range, $1/(1+\eta)$. The line thicknesses indicate one standard deviation of the uncertainty in the estimated values. The three resists were simulated with densities $\rho_\mathrm{CSAR/ZEP}=1.053\si{g}/\si{cm}^3$, $\rho_\mathrm{PMMA}=1.18\si{g}/\si{cm}^3$, and $\rho_\mathrm{HSQ}=0.64\si{g}/\si{cm}^3$ for a number of incident electrons 1e6, 1e5, and 1e5, respectively. The least dense resist, HSQ, shows less scattering at distances shorter than the resist thickness, resulting in a larger $\eta$.}
\end{figure*}

Two additional popular resists, PMMA and HSQ, are compared to that of CSAR (equivalent to ZEP) in Fig.~\ref{fig:appendixPSFresists} and Table~\ref{tab:sim_fit_parameters}, based on simulations of the PPSFs and CDFs at $V=\SI{150}{kV}$. Their primary difference with respect to electron scattering is their densities, which depends significantly on the specific chemical composition. The values used here are $\rho_\mathrm{CSAR/ZEP}=1.053\si{g}/\si{cm}^3$, $\rho_\mathrm{PMMA}=1.18\si{g}/\si{cm}^3$, and $\rho_\mathrm{HSQ}=0.64\si{g}/\si{cm}^3$. The PPSFs and CDFs only deviate slightly: HSQ is lighter and thus has relatively reduced scattering, both in general and at distances shorter than the resist thickness, which results in a larger $\eta$ and less energy deposited per electron.

\begin{table*}[h!]
\centering
\caption{\label{tab:sim_fit_parameters}
Point-spread-function parameters obtained from simulations of electrons impinging on a stack comprising \SI{180}{\nano\meter} resist on silicon at different acceleration voltages, $V$;  Energy deposited per electron, $E_{\mathrm{tot}}/\mathrm{e}^{-}$, and fit parameters of the power-law plus Gaussian (PLG) model, $\beta$, $\eta$, $\gamma$, $\sigma$, and $\nu$.}
\begin{tabular}{cccccccc}
\hline
Resist & $V$ & $E_{\mathrm{tot}}/\mathrm{e}^{-}$ & $\beta$ & $\eta$ & $\gamma$ & $\sigma$ & $\nu$\textsuperscript{a}\\
& \si{\kilo\volt} & $\mathrm{e}$\si{\volt} & \si{\micro\meter} & ~ & \si{nm} & ~ & \si{nm}\\
\hline
CSAR &   5 & 1241 &  0.32(2) & 0.38(4) & 1.68(6) & 2.01(2) & 5.1\\
CSAR &  10 &  617 &  0.81(1.0) & 0.53(6) & 1.86(9) & 2.37(4) & 3.9\\
CSAR &  25 &  242 &  2.57(2) & 0.72(4) & 1.85(5) & 2.58(2) & 3.1\\
CSAR &  50 &  122 &  8.27(9) & 0.72(5) & 1.50(5) & 2.50(2) & 2.9\\
CSAR & 100 &   66 & 27.7(3) & 0.73(4)  & 1.41(4) & 2.45(1.3) & 2.9\\
CSAR & 150 &   51 & 57.2(11) & 0.72(3)  & 1.40(4) & 2.43(1.2) & 2.9\\
PMMA & 150 &   60 & 57.5(20) & 0.73(3)  & 1.40(8) & 2.48(2) & 2.7\\
HSQ & 150 &   25 & 56.7(23) & 0.81(7)  & 1.4(1.0) & 2.42(3) & 3.0\\
\hline
\end{tabular}
\\\textsuperscript{a} $\nu$ values required for normalization with the corresponding $\gamma$ and $\sigma$.
\end{table*}

\vspace{1cm}
\footnotesize
\bibliography{references}

@inproceedings{albrechtsen_power-law_2019,
	title = {{Power-Law Short Range Point-Spread Function In Electron-Beam Lithography}},
	booktitle = {45th {International} {Conference} on {Micro} and {Nano} {Engineering}},
	author = {Albrechtsen, Marcus and Stobbe, Søren},
	year = {2019},
    address = {Rhodes, Greece},
	pages = {PA21},
}

@article{bickford_hydrogen_2014,
	title = {{Hydrogen Silsesquioxane on SOI Proximity and Microloading Effects Correction From a Single 1D Characterization Sample}},
	volume = {32},
	doi = {10.1116/1.4901567},
	number = {6},
	journal = {J. Vac. Sci. Technol. B},
	author = {Bickford, Justin R. and Lopez, Gerald and Belic, Nikola and Hofmann, Ulrich},
	year = {2014},
	pages = {06F511},
}

@phdthesis{manfrinato_electron-beam_2015,
	title = {{Electron-Beam Lithography Towards the Atomic Scale and Applications to Nano-Optics}},
	school = {Massachusetts Institute of Technology},
	author = {Manfrinato, Vitor Riseti},
	year = {2015},
}

@article{gangnaik_new_2017,
	title = {{New Generation Electron Beam Resists: A Review}},
	volume = {29},
	doi = {10.1021/acs.chemmater.6b03483},
	number = {5},
	journal = {Chem. Mater.},
	author = {Gangnaik, Anushka S. and Georgiev, Yordan M. and Holmes, Justin D.},
	year = {2017},
	pages = {1898--1917},
}

@article{manfrinato_determining_2014,
	title = {{Determining the Resolution Limits of Electron-Beam Lithography: Direct Measurement of the Point-Spread Function}},
	volume = {14},
	doi = {10.1021/nl5013773},
	number = {8},
	journal = {Nano Lett.},
	author = {Manfrinato, Vitor R. and Wen, Jianguo and Zhang, Lihua and Yang, Yujia and Hobbs, Richard G. and Baker, Bowen and Su, Dong and Zakharov, Dmitri and Zaluzec, Nestor J. and Miller, Dean J. and Stach, Eric A. and Berggren, Karl K.},
	year = {2014},
	pages = {4406--4412},
}

@article{rishton_point_1987,
	title = {{Point Exposure Distribution Measurements for Proximity Correction in Electron Beam Lithography on a Sub-100 nm Scale}},
	volume = {5},
	doi = {10.1116/1.583847},
	number = {1},
	journal = {J. Vac. Sci. Technol. B},
	author = {Rishton, S. A. and Kern, D. P.},
	year = {1987},
	pages = {135},
}

@article{mankiewich_measurements_1985,
	title = {{Measurements of Electron Range and Scattering in High Voltage E-Beam Lithography}},
	volume = {3},
	doi = {10.1116/1.583204},
	number = {1},
	journal = {J. Vac. Sci. Technol. B},
	author = {Mankiewich, P. M. and Jackel, L. D. and Howard, R. E.},
	year = {1985},
	pages = {174},
}

@article{tennant_progress_2013,
	title = {{Progress and Issues in E-Beam and Other Top Down Nanolithography}},
	volume = {31},
	doi = {10.1116/1.4813761},
	number = {5},
	journal = {J. Vac. Sci. Technol. A},
	author = {Tennant, Donald M.},
	year = {2013},
	pages = {050813},
}

@article{clauset_power-law_2009,
	title = {{Power-Law Distributions in Empirical Data}},
	volume = {51},
	doi = {10.1137/070710111},
	number = {4},
	journal = {SIAM Rev.},
	author = {Clauset, Aaron and Shalizi, Cosma Rohilla and Newman, M. E. J.},
	year = {2009},
	pages = {661--703},
}

@article{owen_methods_1990,
	title = {{Methods for Proximity Effect Correction in Electron Lithography}},
	volume = {8},
	doi = {10.1116/1.585179},
	number = {6},
	journal = {J. Vac. Sci. Technol. B},
	author = {Owen, Geraint},
	year = {1990},
	pages = {1889},
}

@article{saitou_change_1972,
	title = {Change of {Apparent} {Sensitivity} of an {Electron} {Resist} {Due} to {Backing} {Materials}},
	volume = {11},
	doi = {10.1143/JJAP.11.1061},
	number = {7},
	journal = {Jpn. J. Appl. Phys.},
	author = {Saitou, Norio and Munakata, Chusuke and Honda, Yukio},
	year = {1972},
	pages = {1061--1062},
}

@article{saitou_monte_1973,
	title = {Monte {Carlo} {Simulation} for the {Energy} {Dissipation} {Profiles} of 5–20 {keV} {Electrons} in {Layered} {Structures}},
	volume = {12},
	doi = {10.1143/JJAP.12.941},
	number = {6},
	journal = {Jpn. J. Appl. Phys.},
	author = {Saitou, Norio},
	year = {1973},
	pages = {941--942},
}

@inproceedings{p_kern_novel_1980,
	author = {P. Kern, Dieter},
    title = {{A Novel Approach to Proximity Effect Correction}},
	booktitle = {Electron and Ion Beam Science and Technology, Ninth International Conference},
    volume = {80},
	pages = {326-339},
	year = {1980},
}

@article{chang_proximity_1975,
	title = {{Proximity Effect in Electron‐Beam Lithography}},
	volume = {12},
	doi = {10.1116/1.568515},
	number = {6},
	journal = {J. Vac. Sci. Technol.},
	author = {Chang, T. H. P.},
	year = {1975},
	pages = {1271--1275},
}

@article{parikh_selfconsistent_1978,
	title = {{Self‐Consistent Proximity Effect Correction Technique for Resist Exposure ({SPECTRE})}},
	volume = {15},
	doi = {10.1116/1.569678},
	number = {3},
	journal = {J. Vac. Sci. Technol.},
	author = {Parikh, Mihir},
	year = {1978},
	pages = {931--933},
}

@article{parikh_corrections_1979,
	title = {{Corrections to Proximity Effects in Electron Beam Lithography. I. Theory}},
	volume = {50},
	doi = {10.1063/1.326423},
	number = {6},
	journal = {J. Appl. Phys.},
	author = {Parikh, Mihir},
	year = {1979},
	pages = {4371--4377},
}

@article{owen_proximity_1983,
	title = {{Proximity Effect Correction for Electron Beam Lithography by Equalization of Background Dose}},
	volume = {54},
	doi = {10.1063/1.332426},
	number = {6},
	journal = {J. Appl. Phys.},
	author = {Owen, Geraint and Rissman, Paul},
	year = {1983},
	pages = {3573--3581},
}

@article{chou_nanoimprint_1996,
	title = {{Nanoimprint Lithography}},
	volume = {14},
	doi = {10.1116/1.588605},
	number = {6},
	journal = {J. Vac. Sci. Technol. B},
	author = {Chou, Stephen Y. and Krauss, Peter R. and Renstrom, Preston J.},
	year = {1996},
	pages = {4129},
}

@book{nanolithography_2014,
    title = {{Nanolithography}},
	publisher = {Woodhead Publishing},
	pages = {1-41,42-79,80-115,315-347},
	year = {2014},
	author = {B.W. Smith and B.J. Rice and T.R. Groves and D. Resnick},
}

@article{randall_next_2019,
	title = {{Next Generation of Extreme-Resolution Electron Beam Lithography}},
	volume = {37},
	doi = {10.1116/1.5119392},
	number = {6},
	journal = {J. Vac. Sci. Technol. B},
	author = {Randall, John N. and Owen, James H. G. and Lake, Joseph and Fuchs, Ehud},
	year = {2019},
	pages = {061605},
}

@article{jones_very_1987,
	title = {{Very High Voltage (500 kV) Electron Beam Lithography for Thick Resists and High Resolution}},
	volume = {5},
	doi = {10.1116/1.583844},
	number = {1},
	journal = {J. Vac. Sci. Technol. B},
	author = {Jones, G. A. C. and Blythe, S. and Ahmed H.},
	year = {1987},
	pages = {120},
}

@article{pavkovich_proximity_1986,
	title = {{Proximity Effect Correction Calculations by the Integral Equation Approximate Solution Method}},
	volume = {4},
	doi = {10.1116/1.583369},
	number = {1},
	journal = {J. Vac. Sci. Technol. B},
	author = {Pavkovich, J. M.},
	year = {1986},
	pages = {159},
}

@article{murai_fast_1992,
	title = {{Fast Proximity Effect Correction Method Using a Pattern Area Density Map}},
	volume = {10},
	doi = {10.1116/1.585931},
	number = {6},
	journal = {J. Vac. Sci. Technol. B},
	author = {Murai, Fumio and Yoda, Haruo and Okazaki, Shinji and Saitou, Norio and Sakitani, Yoshio},
	year = {1992},
	pages = {3072},
}

@article{eunsung_seo_dose_2000,
	title = {Dose and {Shape} {Modification} {Proximity} {Effect} {Correction} for {Forward}-{Scattering} {Range} {Scale} {Features} in {Electron} {Beam} {Lithography}},
	volume = {39},
	doi = {10.1143/JJAP.39.6827},
	number = {12S},
	journal = {Jpn. J. Appl. Phys.},
	author = {Eunsung Seo, Eunsung Seo and Ohyun Kim, Ohyun Kim},
	year = {2000},
	pages = {6827},
}

@article{lee_proximity_1991,
	title = {{Proximity Effect Correction in Electron-Beam Lithography: A Hierarchical Rule-Based Scheme—PYRAMID}},
	volume = {9},
	doi = {10.1116/1.585367},
	number = {6},
	journal = {J. Vac. Sci. Technol. B},
	author = {Lee, Soo‐Young and Jacob, Joseph C. and Chen, Chung‐Ming and McMillan, Jo A. and MacDonald, Noel C.},
	year = {1991},
	pages = {3048},
}

@article{johnson_program_1989,
	title = {{A Program for Monte Carlo Simulation of Electron Energy Loss in Nanostructures}},
	volume = {7},
	doi = {10.1116/1.584523},
	number = {6},
	journal = {J. Vac. Sci. Technol. B},
	author = {Johnson, S. and MacDonald, N. C.},
	year = {1989},
	pages = {1513},
}

@article{sewell_control_1978,
	title = {{Control of Pattern Dimensions in Electron Lithography}},
	volume = {15},
	doi = {10.1116/1.569677},
	number = {3},
	journal = {J. Vac. Sci. Technol.},
	author = {Sewell, H.},
	year = {1978},
	pages = {927--930},
}

@article{donnelly_plasma_2013,
	title = {{Plasma Etching: Yesterday, Today, and Tomorrow}},
	volume = {31},
	doi = {10.1116/1.4819316},
	number = {5},
	journal = {J. Vac. Sci. Technol. A},
	author = {Donnelly, Vincent M. and Kornblit, Avinoam},
	year = {2013},
	pages = {050825},
}

@article{albrechtsen_nanometer-scale_2022,
	title = {{Nanometer-Scale Photon Confinement in Topology-Optimized Dielectric Cavities}},
	volume = {13},
	doi = {10.1038/s41467-022-33874-w},
	number = {1},
	journal = {Nat. Commun.},
	author = {Albrechtsen, Marcus and Vosoughi Lahijani, Babak and Christiansen, Rasmus Ellebæk and Nguyen, Vy Thi Hoang and Casses, Laura Nevenka and Hansen, Søren Engelberth and Stenger, Nicolas and Sigmund, Ole and Jansen, Henri and Mørk, Jesper and Stobbe, Søren},
	year = {2022},
	pages = {6281},
}

@article{eisenmann_proxeccoproximity_1993,
	title = {{PROXECCO—Proximity Effect Correction by Convolution}},
	volume = {11},
	doi = {10.1116/1.586594},
	number = {6},
	journal = {J. Vac. Sci. Technol. B},
	author = {Eisenmann, Hans and Waas, Thomas and Hartmann, Hans},
	year = {1993},
	pages = {2741},
}

@inproceedings{mack_electron-beam_2001,
    author = {Chris A. Mack},
    title = {{Electron-Beam Lithography Simulation for Mask Making: VI. Comparison of 10- and 50-kV GHOST Proximity Effect Correction}},
    volume = {4409},
    booktitle = {Photomask and Next-Generation Lithography Mask Technology VIII},
    publisher = {SPIE},
    pages = {194 -- 203},
    year = {2001},
    doi = {10.1117/12.438354},
}

@article{hawryluk_energy_1974,
	title = {{Energy Dissipation in a Thin Polymer Film by Electron Beam Scattering}},
	volume = {45},
	doi = {10.1063/1.1663629},
	number = {6},
	journal = {J. Appl. Phys.},
	author = {Hawryluk, R. J. and Hawryluk, Andrew M. and Smith, Henry I.},
	year = {1974},
	pages = {2551--2566},
}

@article{biswas_advances_2012,
	title = {{Advances in Top–Down and Bottom–Up Surface Nanofabrication: Techniques, Applications \& Future Prospects}},
	volume = {170},
	doi = {10.1016/j.cis.2011.11.001},
	number = {1-2},
	journal = {Adv. Colloid Interface Sci.},
	author = {Biswas, Abhijit and Bayer, Ilker S. and Biris, Alexandru S. and Wang, Tao and Dervishi, Enkeleda and Faupel, Franz},
	year = {2012},
	pages = {2--27}
}

@article{asano_metrology_2017,
	title = {{Metrology and Inspection Required for Next Generation Lithography}},
	volume = {56},
	doi = {10.7567/JJAP.56.06GA01},
	number = {6S1},
	journal = {Jpn. J. Appl. Phys.},
	author = {Asano, Masafumi and Yoshikawa, Ryoji and Hirano, Takashi and Abe, Hideaki and Matsuki, Kazuto and Tsuda, Hirotaka and Komori, Motofumi and Ojima, Tomoko and Yonemitsu, Hiroki and Kawamoto, Akiko},
	year = {2017},
	pages = {06GA01}
}

@article{ikeno_electron-beam_2016,
	title = {{Electron-Beam Lithography with Character Projection Technique for High-Throughput Exposure with Line-Edge Quality Control}},
	volume = {15},
	doi = {10.1117/1.JMM.15.3.031606},
	number = {3},
	journal = {J. Micro/Nanolithogr., MEMS, MOEMS},
	author = {Ikeno, Rimon and Maruyama, Satoshi and Mita, Yoshio and Ikeda, Makoto and Asada, Kunihiro},
	year = {2016},
	pages = {031606},
}

@article{servin_progress_2017,
	title = {{Progress and Process Improvements for Multiple Electron-Beam Direct Write}},
	volume = {56},
	doi = {10.7567/JJAP.56.06GC03},
	number = {6S1},
	journal = {Jpn. J. Appl. Phys.},
	author = {Servin, Isabelle and Pourteau, Marie-Line and Pradelles, Jonathan and Essomba, Philippe and Lattard, Ludovic and Brandt, Pieter and Wieland, Marco},
	year = {2017},
	pages = {06GC03},
}

@article{florez_engineering_2022,
	title = {{Engineering Nanoscale Hypersonic Phonon Transport}},
	volume = {17},
	doi = {10.1038/s41565-022-01178-1},
	number = {9},
	journal = {Nat. Nanotechnol.},
	author = {Florez, O. and Arregui, G. and Albrechtsen, M. and Ng, R. C. and Gomis-Bresco, J. and Stobbe, S. and Sotomayor-Torres, C. M. and García, P. D.},
	year = {2022},
	pages = {947--951}
}

@inproceedings{liu_novel_2008,
	title = {{A Novel Curve-Fitting Procedure for Determining Proximity Effect Parameters in Electron Beam Lithography}},
	volume = {7140},
	doi = {10.1117/12.804693},
	booktitle = {Lithography {Asia} 2008},
	author = {Liu, Chun-Hung and Ng, Hoi-Tou and Ng, Philip C. W. and Tsai, Kuen-Yu and Lin, Shy-Jay and Chen, Jeng-Homg},
	year = {2008},
	pages = {367 -- 376}
}

@article{jensen_chemical_1989,
	title = {Chemical {Vapor} {Deposition}},
	volume = {221},
	number = {221},
	journal = {Adv. Chem.},
	author = {Hess, Dennis W. and Jensen, Klavs F.},
	year = {1989},
	doi = {10.1021/ba-1989-0221},
	pages = {199--263},
}

@inproceedings{schirmer_chemical_2013,
    booktitle = {{29th European Mask and Lithography Conference}},
	title = {Chemical Semi-Amplified Positive E-Beam Resist (CSAR 62) for Highest Resolution},
	doi = {10.1117/12.2030576},
	author = {Schirmer, M. and Büttner, B. and Syrowatka, F. and Schmidt, G. and Köpnick, T. and Kaiser, C.},
	year = {2013},
	pages = {88860D},
}

@article{Frase_2009,
doi = {10.1088/0022-3727/42/18/183001},
year = {2009},
volume = {42},
number = {18},
pages = {183001},
author = {Frase, C G and Gnieser, D and Bosse, H},
title = {{Model-Based SEM for Dimensional Metrology Tasks in Semiconductor and Mask Industry}},
journal = {J. Phys. D: Appl. Phys.}
}

@article{yu_systematic_2007,
	title = {{Systematic Considerations for the Patterning of Photonic Crystal Devices by Electron Beam Lithography}},
	volume = {271},
	doi = {10.1016/j.optcom.2006.10.034},
	number = {1},
	journal = {Opt. Commun.},
	author = {Yu, Hejun and Yu, Jinzhong and Sun, Fei and Li, Zhiyong and Chen, Shaowu},
	year = {2007},
	pages = {241--247}
}

@article{abe_proximity_1989,
	title = {{Proximity Effect Correction for High‐Voltage Electron Beam Lithography}},
	volume = {65},
	doi = {10.1063/1.343283},
	number = {11},
	journal = {J. Appl. Phys.},
	author = {Abe, Takayuki and Takigawa, Tadahiro},
	year = {1989},
	pages = {4428--4434},
}

@article{bojko_electron_2011,
    author = {Bojko,Richard J.  and Li,Jing  and He,Li  and Baehr-Jones,Tom  and Hochberg,Michael  and Aida,Yukinori },
    title = {{Electron Beam Lithography Writing Strategies for Low Loss, High Confinement Silicon Optical Waveguides}},
    volume = {29},
	number = {6},
	doi = {10.1116/1.3653266},
	journal = {J. Vac. Sci. Technol. B},
	year = {2011},
	pages = {06F309}
}

@article{perona_scale-space_1990,
	title = {{Scale-Space and Edge Detection Using Anisotropic Diffusion}},
	volume = {12},
	doi = {10.1109/34.56205},
	number = {7},
	journal = {IEEE Trans. Pattern Anal. Mach. Intell.},
	author = {Perona, P. and Malik, J.},
	year = {1990},
	pages = {629-639}
}

@article{ahmad_deposition_2018,
    title = {{Deposition of Nanomaterials: A Crucial Step in Biosensor Fabrication}},
    journal = {Mater. Today Commun.},
    volume = {17},
    pages = {289-321},
    year = {2018},
    doi = {https://doi.org/10.1016/j.mtcomm.2018.09.024},
    author = {Rafiq Ahmad and Otto S. Wolfbeis and Yoon-Bong Hahn and Husam N. Alshareef and Luisa Torsi and Khaled N. Salama},
}

@article{duan_sub10_2010,
    author = {Duan, Huigao and Winston, Donald and Yang, Joel K. W. and Cord, Bryan M. and Manfrinato, Vitor R. and Berggren, Karl K.},
    title = "{{Sub-10-nm Half-Pitch Electron-Beam Lithography by Using Poly(Methyl Methacrylate) As a Negative Resist}}",
    journal = {J. Vac. Sci. Technol. B},
    volume = {28},
    number = {6},
    pages = {C6C58-C6C62},
    year = {2010},
    doi = {10.1116/1.3501353},
}

@article{greve_optimization_2013,
	title = {{Optimization of an Electron Beam Lithography Instrument For Fast, Large Area Writing at 10 {kV} Acceleration Voltage}},
	volume = {31},
	doi = {10.1116/1.4813325},
	pages = {043202},
	number = {4},
	journal = {J. Vac. Sci. Technol. B},
	author = {Greve, Martin M. and Holst, Bodil},
    year = {2013},
}

@article{kanaya_penetration_1972,
	title = {{Penetration and Energy-Loss Theory of Electrons in Solid Targets}},
	volume = {5},
	doi = {10.1088/0022-3727/5/1/308},
	journal = {J. Phys. D: Appl. Phys.},
	author = {Kanaya, K. and Okayama, S.},
	year = {1972},
	pages = {43--58},
}

@article{messina_electron_1992,
	title = {{Electron scattering in microstructure processes}},
	volume = {15},
	doi = {10.1007/BF02742957},
	journal = {Riv. Nuovo Cim.},
	author = {Messina, G. and Paoletti, A. and Santangelo, S. and Tucciarone, A.},
	year = {1992},
	pages = {1--57},
}

@article{adesida_high_1979,
	title = {{High Resolution Electron-Beam Lithography on Thin Films}},
	volume = {16},
	doi = {10.1116/1.570285},
	number = {6},
	journal = {J. Vac. Sci. Technol.},
	author = {Adesida, I. and Everhart, T. E. and Shimizu, R.},
	year = {1979},
	pages = {1743--1748},
}

@article{tennant_electron_1988,
	title = {{Electron Scattering Distribution in InP at 50 kV}},
	volume = {6},
	doi = {10.1116/1.583968},
	number = {1},
	journal = {J. Vac. Sci. Technol. B},
	author = {Tennant, D. M. and Doran, G. E. and Howard, R. E. and Denker, J. S.},
	year = {1988},
	pages = {426--431}
}

@article{wind_proximity_1989,
	title = {{Proximity Correction for Electron Beam Lithography Using a Three-Gaussian Model of the Electron Energy Distribution}},
	volume = {7},
	doi = {10.1116/1.584522},
	number = {6},
	journal = {Journal of Vacuum Science \& Technology B: Microelectronics Processing and Phenomena},
	author = {Wind, S. J. and Rosenfield, M. G. and Pepper, G. and Molzen, W. W. and Gerber, P. D.},
	year = {1989},
	pages = {1507--1512}
}

@article{boere_experimental_1990,
	title = {{Experimental Study on Proximity Effects in High Voltage E-Beam Lithography}},
	volume = {11},
	doi = {10.1016/0167-9317(90)90128-G},
	number = {1-4},
	journal = {Microelectron. Eng.},
	author = {Boere, E. and Van Der Drift, E. and Romijn, J. and Rousseeuw, B.},
	year = {1990},
	pages = {351--354},
}

@article{stevens_determination_1986,
	title = {{Determination of the Proximity Parameters in Electron Beam Lithography Using Doughnut-Structures}},
	volume = {5},
	doi = {10.1016/0167-9317(86)90040-7},
	number = {1-4},
	journal = {Microelectron. Eng.},
	author = {Stevens, L. and Jonckheere, R. and Froyen, E. and Decoutere, S. and Lanneer, D.},
	year = {1986},
	pages = {141--150},
}

@article{dai_experiment-based_2011,
	title = {{Experiment-Based Estimation of Point Spread Function in Electron-Beam Lithography: Forward-Scattering Part}},
	volume = {88},
	doi = {10.1016/j.mee.2011.05.019},
	number = {10},
	journal = {Microelectronic Engineering},
	author = {Dai, Q. and Lee, S.-Y. and Lee, S.-H. and Kim, B.-G. and Cho, H.-K.},
	year = {2011},
	pages = {3054--3061},
}

@article{mcmillan_simulation_1989,
	title = {{Simulation of Electron Beam Exposure of Submicron Patterns}},
	volume = {7},
	doi = {10.1116/1.584529},
	number = {6},
	journal = {J. Vac. Sci. Technol. B},
	author = {McMillan, Jo A. and Johnson, Sylvester and MacDonald, Noel C.},
	year = {1989},
	pages = {1540--1545}
}

@article{liu_impacts_2013,
	title = {{Impacts of Point Spread Function Accuracy on Patterning Prediction and Proximity Effect Correction in Low-Voltage Electron-Beam–Direct-Write Lithography}},
	volume = {31},
	doi = {10.1116/1.4790655},
	number = {2},
	journal = {J. Vac. Sci. Technol. B},
	author = {Liu, Chun-Hung and Ng, Philip C. W. and Shen, Yu-Tian and Chien, Sheng-Wei and Tsai, Kuen-Yu},
	year = {2013},
	pages = {021605},
}

@article{allison_recent_2016,
	title = {{Recent Developments in Geant4}},
	volume = {835},
	doi = {10.1016/j.nima.2016.06.125},
	journal = {Nucl. Instr. Meth. Phys. Res. A},
	author = {Allison, J. and Amako, K. and Apostolakis, J. and Arce, P. and Asai, M. and Aso, T. and Bagli, E. and Bagulya, A. and Banerjee, S. and Barrand, G. and Beck, B.R. and Bogdanov, A.G. and Brandt, D. and Brown, J.M.C. and Burkhardt, H. and Canal, Ph. and Cano-Ott, D. and Chauvie, S. and Cho, K. and Cirrone, G.A.P. and Cooperman, G. and Cortés-Giraldo, M.A. and Cosmo, G. and Cuttone, G. and Depaola, G. and Desorgher, L. and Dong, X. and Dotti, A. and Elvira, V.D. and Folger, G. and Francis, Z. and Galoyan, A. and Garnier, L. and Gayer, M. and Genser, K.L. and Grichine, V.M. and Guatelli, S. and Guèye, P. and Gumplinger, P. and Howard, A.S. and Hřivnáčová, I. and Hwang, S. and Incerti, S. and Ivanchenko, A. and Ivanchenko, V.N. and Jones, F.W. and Jun, S.Y. and Kaitaniemi, P. and Karakatsanis, N. and Karamitros, M. and Kelsey, M. and Kimura, A. and Koi, T. and Kurashige, H. and Lechner, A. and Lee, S.B. and Longo, F. and Maire, M. and Mancusi, D. and Mantero, A. and Mendoza, E. and Morgan, B. and Murakami, K. and Nikitina, T. and Pandola, L. and Paprocki, P. and Perl, J. and Petrović, I. and Pia, M.G. and Pokorski, W. and Quesada, J.M. and Raine, M. and Reis, M.A. and Ribon, A. and Ristić Fira, A. and Romano, F. and Russo, G. and Santin, G. and Sasaki, T. and Sawkey, D. and Shin, J.I. and Strakovsky, I.I. and Taborda, A. and Tanaka, S. and Tomé, B. and Toshito, T. and Tran, H.N. and Truscott, P.R. and Urban, L. and Uzhinsky, V. and Verbeke, J.M. and Verderi, M. and Wendt, B.L. and Wenzel, H. and Wright, D.H. and Wright, D.M. and Yamashita, T. and Yarba, J. and Yoshida, H.},
	year = {2016},
	pages = {186--225},
}

@article{allison_geant4_2006,
	title = {{Geant4 Developments and Applications}},
	volume = {53},
	doi = {10.1109/TNS.2006.869826},
	number = {1},
	journal = {IEEE Trans. Nucl. Sci.},
	author = {Allison, J. and Amako, K. and Apostolakis, J. and Araujo, H. and Arce Dubois, P. and Asai, M. and Barrand, G. and Capra, R. and Chauvie, S. and Chytracek, R. and Cirrone, G.A.P. and Cooperman, G. and Cosmo, G. and Cuttone, G. and Daquino, G.G. and Donszelmann, M. and Dressel, M. and Folger, G. and Foppiano, F. and Generowicz, J. and Grichine, V. and Guatelli, S. and Gumplinger, P. and Heikkinen, A. and Hrivnacova, I. and Howard, A. and Incerti, S. and Ivanchenko, V. and Johnson, T. and Jones, F. and Koi, T. and Kokoulin, R. and Kossov, M. and Kurashige, H. and Lara, V. and Larsson, S. and Lei, F. and Link, O. and Longo, F. and Maire, M. and Mantero, A. and Mascialino, B. and McLaren, I. and Mendez Lorenzo, P. and Minamimoto, K. and Murakami, K. and Nieminen, P. and Pandola, L. and Parlati, S. and Peralta, L. and Perl, J. and Pfeiffer, A. and Pia, M.G. and Ribon, A. and Rodrigues, P. and Russo, G. and Sadilov, S. and Santin, G. and Sasaki, T. and Smith, D. and Starkov, N. and Tanaka, S. and Tcherniaev, E. and Tome, B. and Trindade, A. and Truscott, P. and Urban, L. and Verderi, M. and Walkden, A. and Wellisch, J.P. and Williams, D.C. and Wright, D. and Yoshida, H.},
	year = {2006},
	pages = {270--278},
}

@article{agostinelli_geant4simulation_2003,
	title = {{Geant4—a Simulation Toolkit}},
	volume = {506},
	doi = {10.1016/S0168-9002(03)01368-8},
	number = {3},
	journal = {Nucl. Instr. Meth. Phys. Res. A},
	author = {Agostinelli, S. and Allison, J. and Amako, K. and Apostolakis, J. and Araujo, H. and Arce, P. and Asai, M. and Axen, D. and Banerjee, S. and Barrand, G. and Behner, F. and Bellagamba, L. and Boudreau, J. and Broglia, L. and Brunengo, A. and Burkhardt, H. and Chauvie, S. and Chuma, J. and Chytracek, R. and Cooperman, G. and Cosmo, G. and Degtyarenko, P. and Dell'Acqua, A. and Depaola, G. and Dietrich, D. and Enami, R. and Feliciello, A. and Ferguson, C. and Fesefeldt, H. and Folger, G. and Foppiano, F. and Forti, A. and Garelli, S. and Giani, S. and Giannitrapani, R. and Gibin, D. and Gómez Cadenas, J.J. and González, I. and Gracia Abril, G. and Greeniaus, G. and Greiner, W. and Grichine, V. and Grossheim, A. and Guatelli, S. and Gumplinger, P. and Hamatsu, R. and Hashimoto, K. and Hasui, H. and Heikkinen, A. and Howard, A. and Ivanchenko, V. and Johnson, A. and Jones, F.W. and Kallenbach, J. and Kanaya, N. and Kawabata, M. and Kawabata, Y. and Kawaguti, M. and Kelner, S. and Kent, P. and Kimura, A. and Kodama, T. and Kokoulin, R. and Kossov, M. and Kurashige, H. and Lamanna, E. and Lampén, T. and Lara, V. and Lefebure, V. and Lei, F. and Liendl, M. and Lockman, W. and Longo, F. and Magni, S. and Maire, M. and Medernach, E. and Minamimoto, K. and Mora de Freitas, P. and Morita, Y. and Murakami, K. and Nagamatu, M. and Nartallo, R. and Nieminen, P. and Nishimura, T. and Ohtsubo, K. and Okamura, M. and O'Neale, S. and Oohata, Y. and Paech, K. and Perl, J. and Pfeiffer, A. and Pia, M.G. and Ranjard, F. and Rybin, A. and Sadilov, S. and Di Salvo, E. and Santin, G. and Sasaki, T. and Savvas, N. and Sawada, Y. and Scherer, S. and Sei, S. and Sirotenko, V. and Smith, D. and Starkov, N. and Stoecker, H. and Sulkimo, J. and Takahata, M. and Tanaka, S. and Tcherniaev, E. and Safai Tehrani, E. and Tropeano, M. and Truscott, P. and Uno, H. and Urban, L. and Urban, P. and Verderi, M. and Walkden, A. and Wander, W. and Weber, H. and Wellisch, J.P. and Wenaus, T. and Williams, D.C. and Wright, D. and Yamada, T. and Yoshida, H. and Zschiesche, D.},
	year = {2003},
	pages = {250--303},
}

@article{Manfrinato_Sub-5keV_2011,
title = {{Sub-5keV electron-beam lithography in hydrogen silsesquioxane resist}},
journal = {Microelectron. Eng.},
volume = {88},
number = {10},
pages = {3070-3074},
year = {2011},
doi = {https://doi.org/10.1016/j.mee.2011.05.024},
author = {Vitor R. Manfrinato and Lin Lee Cheong and Huigao Duan and Donald Winston and Henry I. Smith and Karl K. Berggren},
}

@article{Winston_modeling_2012,
  title={{Modeling the point-spread function in helium-ion lithography}},
  author={Winston, Donald and Ferrera, J and Battistella, L and Vlad{\'a}r, AE and Berggren, Karl K},
  journal={Scanning},
  volume={34},
  number={2},
  pages={121--128},
  year={2012},
}

@article{mao_proximity_2025,
  title={{Proximity effect correction in electron beam lithography using a composite function model of electron scattering energy distribution}},
  author={Mao, Qingyuan and Zhu, Jingyuan and Cheng, Xinbin and Wang, Zhanshan},
  journal={Discov. Nano.},
  volume={20},
  number={1},
  pages={84},
  year={2025},
  publisher={Springer}
}

\end{document}